\documentclass[preprint]{aastex}

\shorttitle{Bright contact binary stars}
\shortauthors{Rucinski}

\begin{document}

\title{The 7.5 Magnitude Limit Sample of Bright Short-Period 
Binary Stars. I.\\
How Many Contact Binaries Are There?\footnote{Based on 
data from the Hipparcos satellite mission and from the David Dunlap
Observatory, University of Toronto.}}

\author{Slavek M. Rucinski}
\affil{David Dunlap Observatory, University of Toronto\\
P.O.Box 360, Richmond Hill, Ontario, Canada L4C 4Y6}
\email{rucinski@astro.utoronto.ca}

%\centerline{\today}

\begin{abstract}

A sample of bright contact binary stars (W~UMa-type or EW,
and related: with $\beta$~Lyr light curves, EB, and ellipsoidal, ELL
-- in effect, all but the detached, EA), 
to the limit of $V^{max} = 7.5$ magnitude is deemed
to include all discoverable short-period ($P < 1$ days) binaries 
with photometric variation larger than about 0.05 magnitude.
Of the 32 systems in the final sample, 11 systems have been discovered by 
the Hipparcos satellite. The combined spatial density is evaluated 
at $(1.02 \pm 0.24) \times 10^{-5}$ pc$^{-3}$. 
The Relative Frequency of Occurrence (RFO), defined in relation to
the Main Sequence stars, depends on the
luminosity. An assumption of RFO $\simeq$ 1/500 
for $M_V > +1.5$ is consistent with the data, although
the number statistics is poor with the resulting uncertainty 
in the spatial density and the RFO by a factor of about two.  
The RFO rapidly decreases for brighter binaries to a level of
1/5,000 for $M_V < +1.5$ and to 1/30,000 for $M_V < +0.5$. 
The high RFO of 1/130, previously determined from 
the deep OGLE--I sample of Disk Population
W~UMa-type systems towards Baade's Window,
is inconsistent with and unconfirmed by the new results.
Possible reasons for the large discrepancy are discussed. They include
several observational effects, but also a possibility of a genuine
increase in the contact-binary density in the central parts of the
Galaxy.

\end{abstract}

\keywords{ stars: close binaries - stars: eclipsing binaries -- 
stars: variable stars}

\section{THE CHANGING VIEWS ON THE SPATIAL DENSITY OF THE W UMA SYSTEMS}
\label{intro}

Our views about the spatial density of contact, W~UMa-type
binaries have meandered considerably during the recent half century.
\citet{sha48} was the first to point out that ``W~Ursae Majoris variables ...
are not only the most numerous of eclipsing stars ... but ... more
numerous than all other variable stars combined''. From the numbers
given by Shapley, one can infer that he estimated that perhaps as
many as about one percent of all solar-type stars are W~UMa-type variables. 

Several attempts have been made to put the statement of Shapley
in a quantitative way, with diverse results. \citet{pop64}
estimated the local density at $2 \times 10^{-5}$ pc$^{-3}$, while
\citet{kra67} estimated it to be some 20 times lower,
$10^{-6}$ pc$^{-3}$. Then \citet{vnV75} found a much larger number,
$11 \times 10^{-5}$ pc$^{-3}$, which corresponds to about one percent
of all stars being W~UMa-type binaries. Subsequently 
\citet{due84} derived a value of
$\simeq 10^{-5}$ pc$^{-3}$, equivalent to the Relative Frequency of
Occurrence (RFO) of $\simeq 0.001$ (or 0.1 percent), and this 
value remained
in popular use for some time. A similar, simplified, and thus rather
convincing (based on naked eye objects) 
estimate of the frequency was given by \citet{ruc93},
RFO~$\simeq 0.0005 - 0.002$, that is one W~UMa binary 
per 500 -- 2000 ordinary stars.

The newest investigations, based on much larger statistical
samples, suggested that the RFO is perhaps as high
as suggested by Shapley, i.e.\ some five to ten times higher 
than estimated by \citet{due84}. During the last two decades, 
several EW systems have been discovered in old open clusters
\citep{KR93,RK94,ruc98b}. A combined approximate estimate 
of \citet{ruc94} for a few old open clusters gave the RFO of
$\simeq 275 \pm 75$ ordinary dwarfs per 
one EW system. Later, in a combined analysis of several open
clusters, \citet{ruc98b} showed that the RFO evolves with the
age of the stellar system; the number of EW systems increases 
from the level of one per a thousand at the age of about 0.8 Gyr
to the level of one per some 200 -- 300 dwarfs at the age of 
5 -- 7 Gyr.
The RFO was found also very high in globular clusters
\citet{ruc00}; however, Population~II contact binaries are of
no relevance to the present study which concentrates on the
Population~I objects, primarily in the solar neighborhood.

The highest RFO was estimated for the Galactic Disk, as probed by
the narrow, deep pencil beam of the OGLE--I survey in the
direction of Baade's Window
\citep{ruc95,ruc97a,ruc97b,ruc98a,ruc98b}. This finding was
in a basic accordance with the view of the increased RFO
from the age of the oldest open clusters to the age of the
Galactic Disk Field of about 11 Gyr \citep{bin00}.
The great advantage of the OGLE--I result over all previous 
spatial density estimates was not only in the size and
uniformity of the sample (a few hundred contact systems discovered 
in the same survey), but -- mainly -- in that the RFO
estimate was based on the {\it volume-limited samples\/}, complete
to $M_V \simeq 5.5$ to the distance of 3 kpc, and (with better
statistics for brighter systems) to $M_V \simeq 4.2$ to the 
distance of 5 kpc. Large numbers of binaries in the OGLE--I sample
permitted to evaluate the RFO on the per-$M_V$-bin way,
in place of the previous estimates based on
data averaged for all accessible spectral types, 
between late A to early K. Such
comparisons through the luminosity function led to a statistically 
well established -- and high -- apparent frequency at the level of 
$\simeq 1/130$ \citep{ruc98b}. A clear drop in numbers
for brighter, longer-period contact systems, with a sharp
cut-off at $P \simeq 1.3-1.5$ days, was also noted \citep{ruc98a,
ruc98b}.

The OGLE--I result would suggest a return
to Shapley's value of the RFO at the level of one percent or higher.
However, this would mean a rather large disagreement with the
relatively firm result of \citet{due84} for the solar
neighborhood. Can we
exclude a systematic error in the OGLE--I result? Even
if based on the best {\it number statistics\/}, 
the OGLE--I may have {\it systematic\/} biases.
After finding that the new 7.5 magnitude-limit sample indeed
does not agree with the OGLE--I result, we attempt to
give possible reasons for the discrepancy in Section~\ref{discus}.

Sections~\ref{criteria} -- \ref{PC}
describe construction and properties 
of the sample used in a new estimate of 
the spatial density. This estimate is through the luminosity
function, as described in Section~\ref{LF}. The related 
``period function'' (the number of systems per unit
of volume, per period interval) is described in Section~\ref{PF}.
The amplitude distribution for the sample is discussed in
Section~\ref{ampl}. Section~\ref{discus} 
discusses the major discrepancy in the density
spatial estimates between the current sample and the 
OGLE--I sample in the direction of Baade's Window \citep{ruc98b}. 
Section~\ref{future} looks into
the future of the spatial density estimates for contact binaries.
Section~\ref{concl} summarizes the results of the paper.
Appendix~\ref{appendix} gives brief descriptions of the
individual binaries, including those which have been
excluded from the sample in the last stages of its definition.

\section{CRITERIA FOR SELECTION OF THE 7.5 MAGNITUDE SAMPLE}
\label{criteria}

A well defined volume-limited sample would be the
proper way of estimating the spatial density of the EW systems.
The Hipparcos sample immediately comes to mind because --
in principle -- the parallax
data can be directly used to place systems in space and thus lead
to straightforward estimates of the
numbers per unit of volume. The Hipparcos sample of bright stars
is also currently the
most complete as far as stellar variability is concerned, down to
the amplitude levels perhaps 0.02 -- 0.03 magnitude, and 
certainly to amplitudes larger than 0.05 magnitude. As we
will see, in the final sample that we considered of 32 close binary
systems, 11 systems are new Hipparcos discoveries, so
that the increase in numbers is significant and must be
considered in the spatial density estimates. However, the completeness
of the Hipparcos sample is in fact limited to the bright stars, 
to $V \simeq 7.5$. The much deeper Hipparcos sample of the W~UMa-type 
systems discussed by \citet{RD97} was {\it not complete\/} in any sense
as it relied on the input lists of all then known systems which was
provided to the Hipparcos project when
the mission was programmed. Thus the list
of Rucinski \& Duerbeck cannot be used to evaluate the space density
of the W~UMa-type systems. 

Of the two advantages of the Hipparcos data for bright stars, the
availability of parallaxes and the completeness for photometric
variability detection, the latter appears -- paradoxically -- to be
more important. As is well known, Hipparcos obtained the first set
of reliable trigonometric parallaxes. But they still had relatively large
random errors, as can be illustrated by the statistics of parallaxes 
for our final sample. While the maximum value
of the parallax (44~Boo) is 78.39 mas (milliarcsec), 
the median value for the sample is only 
$\overline{\pi} = 12.40$ mas, with the smallest value being 4.37 mas.
This should be compared with the median of the mean standard errors 
$\overline{\sigma (\pi)} = 0.85$, with the smallest and largest errors being
0.41 and 2.36 mas\footnote{Four binaries with the 
parallax errors larger than 2 mas, HT~Vir, V867~Ara, DX~Dra and AA~Cet,
are members of triple systems. The three former have been eliminated  
from the sample for other reasons so that only AA~Cet remains in the
sample. More about AA~Cet in Section~\ref{PC}, where the 
very low value of the parallax and its large error
are specifically addressed.}. 
Recalling that, as was shown by \citet{lut73},
large {\it systematic\/} errors in distances start occurring 
for parallaxes smaller than $\pi \simeq 7 \, \sigma (\pi)$, we cannot use
the parallax data to {\it select\/} the sample. 

Thus, knowing that construction of a {\it volume-limited\/}
sample still remains impossible, 
we try to construct a complete {\it magnitude-limited\/}
sample and then use it to obtain the spatial density estimates
in intervals of $M_V$. With this goal in mind, we constructed
the 7.5 magnitude-limit sample, in the following way:
We selected the short-period binaries by merging data from the
two basic sources, the Hipparcos Catalog \citep[HIP]{hip} and the 
General Catalogue of Variable Stars \citep[GCVS]{gcvs4}. The
selected binary stars had the orbital period shorter than
one day and were brighter than 7.5 magnitude at maximum light.
Since the number of systems rapidly increases with the
limiting magnitude, the magnitude cutoff was particularly
carefully observed, as described in Section~\ref{magnitudes}. The 
version of the GCVS which was used is available only by the
Internet\footnote{http://www.sai.msu.su/groups/cluster/gcvs/gcvs/iii/}
and consists of Volumes I -- III of the Fourth Edition of 1988 merged
with the Name Lists Nos. 67 -- 74. We added to this the 
Name List 75 \citep{kaz00}. This material was considered complete
in the Fall of 2001. To limit the sample to the nearby stars,
we removed stars with the HIP parallaxes smaller than 
2.5 milli-arcsec (mas), i.e.\ more distant than 400 pc, 
and, thus -- with our magnitude limit -- brighter 
in absolute sense than $M_V = -0.5$. 

The sample, at this stage consisting of 113 variable stars, 
underwent further scrutiny for presence of pulsating stars, spurious
variables or variables with no orbital periods (formally zero days),
but with single minima (probably detached eclipsing binaries)
observed by Hipparcos. All relevant literature information was used, 
particularly on $\delta$~Sct and $\gamma$~Dor pulsating stars,
leading to a reduction in the sample to 46 stars. Of particular
interest are the following items:
\begin{itemize}
\item Three variables are included in the HIP Catalog as EB systems, 
but appear to be pulsating variables. This result is based on 
our own spectroscopic data obtained at the David Dunlap Observatory
(Rucinski et al., private communication).
These are the small amplitude (0.04 -- 0.05 mag.) variables: CC~Lyn 
(HD~60335, HIP~36965, $V^{max} = 6.40$), 
FH~Cam (HD~63383, HIP~38900, $V^{max} = 6.88$) 
and CU~CVn (HD~120349, HIP~67357, $V^{max} = 7.53$). 
The two former show sharp-line spectra
with radial-velocity changes within 10 to 20 km~s$^{-1}$ and
8 to 14 km~s$^{-1}$, while the third (whose radial
velocity variations remain to be analyzed) shows rotationally
broadened spectra with $V~\sin i = 155 \pm 10$ km~s$^{-1}$.
In all three 
cases the periods appear to be half of the HIP periods, i.e.\
0.177311, 0.13624 and 0.067834 days. 
Note that without spectroscopic data these three stars would be 
classified and included by us among the
ellipsoidal binaries (ELL). The final sample of ELL variables
consists of 14 objects so that removal of the three stars was
significant; these numbers can be also taken as an 
estimate of the possible level of the contamination of the sample 
by the pulsating variables. 
\item While we excluded all detached binaries from our 
sample, we scrutinized all which fell within our period and
brightness limits in order not to miss relevant stars which 
could be incorrectly classified. Among 
the five detached binaries in the HIP catalog with EA designation,
fulfilling our brightness and period criteria, 
three binaries: MX~Hya (0.7960 d), UU~Psc (0.8417 d) and V772~Her 
(0.8795 d), do not have reliable HIP light curves, and only
two have reasonable HIP light curves: BB~Scl (0.4765 d) and 
ER~Vul (0.6981 d). We analyzed the latter together with 
the contact binaries, mostly as a check of our procedures.
It should be noted 
that ER~Vul frequently appears in various catalogues under an
incorrect entry of EW (we expand on the naming conventions
EW, EB, ELL and EA in Section~\ref{vartypes}).
\item We excluded two, most-likely EA binaries from consideration:
BR~Ind and V2300~Oph. 
While BR~Ind appears in the HIP Catalog with the period
of 0.8928 day, its period is most likely twice that long. The same 
possibility applies to the extremely small amplitude variable V2300~Oph 
\citep{jer93} with the nominal period of 0.9071 day. 
\end{itemize}

\section{$\mathbf{H_P}$ AND $\mathbf{V}$ 
MAGNITUDES AND THE 7.5 MAGNITUDE LIMIT}
\label{magnitudes}

Our selection of the sample and assignment of variability types
was based on the Hipparcos light curves in the instrumental
$H_P$ system. Since the number of objects quickly increases with
the limiting magnitude, it is important to note that the cutoff 
at the maximum-light 
7.5 magnitude is in the standard $V$ magnitude bandpass.
For each Hipparcos light curve, 
the maximum $H_P$ magnitude was determined. The values are given in 
in individual panels of Figures~\ref{fig1} and \ref{fig2}. 
The order in these figures is the same as in 
Table~\ref{tab1} and follows the increasing orbital period. 

The values of $H_P^{max}$ have been transformed into $V^{max}$
using the table of the differences $V-H_P$ as a function $V-I$, 
as given in Table~1.3.5.\ in the HIP Catalog Explanatory Books. 
The $V-I$ indices were in turn estimated from the mean
$B-V$ values, as derived from the Tycho $B_T$ and $V_T$ data
in the Tycho-2 catalogue \citep[TYC2]{tyc2}, using the appropriate
linear transformations.
Since most of the objects do not have ground-based photometric indices,
this source of the average $B-V$ data assures some uniformity for
inter-comparison. However, the resulting values of $V^{max}$ 
may be uncertain because of the two-step interpolation of $V-H_P$. 
For that
reason, whenever available, ground-based values of $V^{max}$ have
been used in preference to the values determined from $H_P^{max}$.
Two EW systems, V759~Cen and V566~Oph, have $H_P^{max}$
below the cutoff limit, but their transformed place them 
at $V^{max} < 7.5$, confirming the available literature data.

Table~\ref{tab1} 
contains the relevant data for the systems of the sample.
Notes on individual objects are given in the Appendix~\ref{appendix}.
Of the 43 objects listed in the table, 
two (BB~Scl and ER~Vul, listed at the end of the table) 
were known from the beginning to be EA systems. 
Six EA systems were later identified among binaries originally 
classified as EW and EB (the variability-type classification is
discussed in full in Sections~\ref{vartypes} and
\ref{Fourier}). Finally, three 
W~UMa-type systems in triple systems have been rejected 
because only the combined brightness
with their close companions was above the adopted magnitude limit; 
this is discussed in the next Section~\ref{thirdlight}. 

The final sample of contact and related stars (EW + EB + ELL)
consists of 32 objects. The sample does not contain 
many popular systems repeatedly
observed over the years, among them the prototype of the class,
W~UMa. On the other hand, Table~\ref{tab1} 
contains 17 new Hipparcos discoveries,
of which 11 have survived the selection scrutiny to be
included in the final sample. These numbers best 
illustrate that a re-discussion of the current statistics, 
even for a magnitude-limited sample with a bright limit, 
was very much needed.

\section{THE THIRD LIGHT: CLOSE COMPANIONS}
\label{thirdlight}

Several among the binaries of the sample have  
speckle interferometry, close visual or spectroscopic companions.
The light contribution of a third component will be characterized
here by the fractional light factor, $\beta = L_3/L_{12}^{max}$,
the ratio of the observed luminosity of the third component
to the observed luminosity of the close binary at its maximum
brightness. 
%For components of equal brightness, $\beta = 1$. 
The values of $\beta$ for individual cases were found from the
literature, as given in the Appendix, usually after 
consultation of the very useful studies of \citet{tok97} 
and \citet{FM00}. 
We list the adopted values of $\beta$ in Table~\ref{tab1}. 
We also give there comments on the angular separation (in arcsec). 
The final, corrected values of the maximum magnitude for the binary
systems $V^{max}$ have been computed using: 
$V^{max}(corr) = V^{max}(obs) + 2.5 \log (1+\beta)$. 

The current sample has been reduced in size by taking into account 
the presence of close ``third components'' in  
systems which -- without allowance for the companions --
were brighter than the 7.5 magnitude, but which became fainter
than this limit when the ``third light'' was subtracted. 
The three systems, included in Table~\ref{tab1}, but 
excluded from any further statistical consideration  
were HT~Vir, KR Com and V867~Ara. The downward shifts
from the observed values of $V^{max}$ to the corrected ones
were moderately small (0.41, 0.16, 0.24 mag.), 
but significantly large to cross the 7.5 magnitude limit. We
are particularly sorry to part with the system HT~Vir which is
one of the most interesting in the contact-binary world: It shows
a large amplitude of light variations (when corrected for the third
light), in perfect agreement with its large mass ratio,
$q_{sp}=0.812 \pm 0.008$ \citep{ddo4}. 

Presence of a third light in some system leads to a ``dilution'' of
the variability signal in the light curve with the corresponding
decrease in size of the Fourier coefficients used for variability 
classification (see Section~\ref{Fourier}). To account for this effect,
the light curve decomposition coefficients, expressed in light units,
have been corrected for the third-light 
using: $a_i(corr) = a_i(obs) (1+\beta)$.
The light variation amplitudes ($amp$), measured in magnitudes, also
change according to: $amp (corr) = -2.5 \log [(1+\beta) \>
{\rm dex}(-0.4 \> amp (obs)) - \beta]$.

\section{VARIABILITY TYPES}
\label{vartypes}

The goal of this paper is 
to evaluate the spatial density of contact binary stars.
Three classes of binaries must be considered in this context, the
EW or W~UMa-type, the EB or Beta Lyrae-type binaries
and the ELL or Ellipsoidal variables. 
The distinction between them is sometimes difficult, especially when
color curves are not available.

The EW or classical contact or W~UMa-type
binaries are characterized by continuous light
variations and equally deep eclipses. The latter property results from
equal temperatures of components, in spite of their differing masses.
The equality of the temperature and thus very good thermal contact
can be taken as the best defining characteristics of the contact-binary
class. Normally, EW binaries show a slight reddening at both eclipses
due to the combined effects of the gravity and limb-darkening 
which dominate over a small temperature difference between
the components.

Observationally, the class EB is defined as basically a modification
of the EW class, but with unequally deep eclipses, leading to
light curves of the $\beta$ Lyrae-type. 
Since the differing depth of the minima is 
a component temperature-difference effect, 
the color curves show reddening and
blueing in primary and secondary minima, respectively. Spectroscopists
tend to call EB systems those that show lines of only one component.
Physically, what is observed as the EB binaries is, unfortunately, a 
mixture of distinct classes of (1) contact binaries in poor 
thermal contact, (2) pre-contact or broken-contact 
semi-detached binaries and (3) semi-detached
Algols, after the mass-exchange and mass-ratio reversal. 
Thus, the classification of EB may be occasionally misleading. 
The real, massive and evolved,
$\beta$~Lyrae-type binaries (sometimes also called
W~Ser binaries), with the inverted mass-ratio
and on-going accretion into a large accretion disk,
do not exist among very short-period stars (there is no space
for the disk).  
There are also indications that for periods shorter than one day, 
Algols are extremely rare \citep{WCrv}. The most frequently occurring 
short-period EB objects appear to be binaries either 
just before establishing contact or in one
of the broken-contact semi-detached stage, 
in any case, prior to the mass-ratio reversal, with the more
massive component close to or at its critical Roche surface.

The distinction between the EW and EB class in individual cases
is frequently difficult even when radial velocity data for
both components are available. In our radial-velocity program 
conducted at the David Dunlap Observatory, we have observed systems 
which have been originally called EB on the basis that only
one set of spectral lines had been observed in previous programs, 
but for which we see now spectra of both stars, sometimes suggesting
a different classification. Frequently, no directly available data 
%-- except perhaps after extensive modeling -- 
can tell us if the star is an
EW system with a broken thermal contact or an EB system with 
the more massive star filling the Roche lobe and an undersized secondary.
In both cases, the more massive component is the hotter one. 
Both types are evolutionary related with one preceding or leading to
the other. Thus, we decided to use both classes, EW and EB, in a 
combined spatial density estimate. We note that \citet{ste01}
(also St\c{e}pie\'n 2001, private communication)
suggested that contact binaries are basically semi-detached objects
with the smaller component engulfed by the matter of the more
massive, larger component. This presumably dynamically-stable
modification of the Shu model \citep{SL81}
still requires an actual modeling effort. If this model would work,
the distinction between the EW and EB classes would reduce only to the
efficiency of the energy transport. 

The third class are the ellipsoidal, or ELL, variables. 
This class has no real physical meaning, as these 
are -- in majority -- EW and EB binaries seen at
small orbital inclination angles.
We simply have no way of assigning these low-amplitude variables to 
the two classes. Predictions on the numbers of low amplitude
contact systems \citep{ruc01} suggest that there should be
many such variables and indeed the Hipparcos mission discovered
majority of them among the bright stars. As discussed in \citet{ruc01}, 
some ELL variables may be
very low mass-ratio objects seen at large inclinations, but then
total eclipses can be observed even for relatively low inclinations.

Both catalogs used in the selection of the current sample,
the HIP Catalog and the Variable Star Catalog,
give the variability types. However, 
we found that these are frequently unreliable. 
This was particularly so for the large number of small-amplitude 
systems called EB in the HIP Catalog,
most probably by some sort of an automatic classifier. Of
course, these could be in fact EW or EB binaries, but 
the tendency to call all variables EB seemed to increase 
for very small amplitudes, exactly in the regime where 
usually nothing can be said solely on the basis of the light curve. 
For that reason, we used the name of ELL variables to
signify all small amplitude systems with amplitudes smaller
than 0.1 magnitude (more strictly, when $|a_2| < 0.05$).

Our goal was to determine the density of contact binaries which
we here identify with the combined EW + EB + ELL classes.
The detached (EA) binaries have been eliminated early from
consideration, mostly because the 
statistics of short-period detached binaries is practically 
impossible to establish at this time on the basis of
the photometric data alone. Also, while reasonably secure
predictions on the number of systems missed due to
low inclinations can be made for stars filling the Roche lobes, 
mostly because of the relatively simple geometry
\citep{ruc01}, such predictions are 
entirely impossible for detached binaries for which the
parameter space is too large. In the final
sample, we identified and eliminated the EA binaries
through the Fourier analysis of the Hipparcos
light curves, as described in Section~\ref{Fourier}.

\section{FOURIER ANALISIS OF THE LIGHT CURVES AND VARIABILITY
CLASSIFICATION}
\label{Fourier}

The light curves obtained by the 
Hipparcos mission in the $H_P$ wide--band photometry were
retrieved from the HIP database for all 
43 close binaries with periods shorter than one day
and the maximum brightness above the $V=7.5$ magnitude limit.
We show the light curves in Figures \ref{fig1} and \ref{fig2}.

The classification and selection of the objects has been done
by performing the Fourier analysis of the
Hipparcos light curves in the same way  
as described originally in \citet{ruc97a,ruc97b}. 
The Fourier coefficients are listed in Table~\ref{tab1}. They
have been determined by least-squares fits
to light curves expressed in light units (using the assumed
$H_P^{max}$ as the reference level),
with errors estimated through a ``bootstrap'' re-sampling process.
The median value of the mean standard error 
for all coefficients was 0.0018 (this value is identical for
all coefficients for a given light curve 
because of the orthogonality), but
with some errors -- for poorly-covered light curves -- reaching
much larger values than the sample median. 
Note in particular that $a_1$, which by
definition should be negative, is positive for some systems. The
positive values of $a_1$ 
are invariably associated with large errors due to poor
light curves or incomplete phase coverage, e.g.\ for V759~Cen
and DX~Aqr the mean standard errors of $a_1$ are 0.0045 and 0.0055.
For some stars, the coefficients and their errors
have been magnified when allowance for the third light was made
(Section~\ref{thirdlight}). Such corrected
coefficients are shown in the figures described below; they can
be eaily identified by the unusually large error bars in the
figures.

The separation of the EW and EB systems from the detached binaries, EA,
was done in the $a_2 - a_4$ plane. The ``contact'' line in 
Figure~\ref{fig3} cleanly separates the detached binaries 
from the EW/EB systems which are located below this line.
The ELL systems in this plane are these binaries which are not 
obviously EA and show small $a_2$ (in the absolute sense),
$-0.05 < a_2 < 0$; this approximately corresponds to the variability
amplitude $<0.1$ magnitude. 
Six EA binaries have been identified and removed from
the sample at this stage:  
VW~Pic, V449~Aur, V1130~Tau and GK~Cep appear to consist of 
strongly tidally-distorted, detached components of similar
brightness, while HL~Dra and DX~Aqr, appear to be detached, 
but with components of different temperatures (see below).

Figure~\ref{fig4}
shows the $a_2 - a_1$ coefficient plane which can be
used to separate the EW and EB systems. The first cosine term, $a_1$, is
sensitive to the temperature difference between the components. 
We used the criterion of $a_1 < -0.02$, the same as in \citet{ruc97b},
to assign systems to the EB class. There are only five 
EB systems in the sample: KP~Peg, FO~Vir, ES~Lib, AA~Cet and V1010~Oph. 
The other systems above the $a_1 = -0.02$ line, DX~Aqr and HL~Dra,
are the two EA binaries with components of different temperatures.

The third of the Fourier-decomposition diagrams
(Figure~\ref{fig5}) addresses the light-curve asymmetry, measured
by the first sine term, $b_1$. It is clear that the O'Connell effect
\citep{DM84}, of the first light-curve maximum being higher, 
is clearly present in V1010~Oph
and ES~Lib. The type of the correlation between $a_1$ and $b_1$
was quite prominently visible in the OGLE--I data 
for EB binaries \citep{ruc97b}. It was interpreted 
as a result of accretion phenomena in a semi-detached 
configuration with the more massive star filling the Roche.

\section{THE PERIOD--COLOR RELATION AND THE ABSOLUTE MAGNITUDE
CALIBRATION}
\label{PC}

The period--color relation is shown
in Figure~\ref{fig6}. This relation may be useful
in pointing systems which are evolved and/or reddened as the two
effects move stars away from the Short-Period Blue Envelope (SPBE)
which has a meaning somewhat similar to the Zero-Age Main Sequence
line in the color--magnitude diagrams. It may also uncover 
inconsistencies and misclassifications, such as inclusion of
pulsating stars. The approximating relation
shown for the SPBE, $(B-V)_{SPBE} = 0.04 \> P^{-2.25}$, with the
period in days \citep{ruc98b}, does not have any physical significance,
but is used for simplicity. We used the observed $B-V$ color
indices, without any corrections for the interstellar reddening, which
should be small for most of the nearby objects of the sample. 

The $B-V$ indices used in Figure~\ref{fig6}
have been corrected for the third-light contamination only for 44~Boo
where the brighter companion dominates all observed characteristics
of the close binary; $(B-V)_B=0.94$ \citep{FM00}. 
We were unable to evaluate corrections
to the color indices for the five multiple systems which 
remained in the final sample. For VW~Cep, KP~Peg, V2388~Oph and ES~Lib,
the light contribution of the third body is moderately 
small ($\beta \le 0.22$), while for AA~Cet, 
with $\beta=0.59$, we have no information on the color 
index of the companion, but its spectral type is similar 
to that of the binary \citep{tok97}.

As can be seen in Figure~\ref{fig6}, two objects have abnormally
blue color indices, V445~Cep and KS~Peg. With no spectroscopic data
currently available for these stars, these peculiar colors have no
explanation. We cannot exclude a possibility that these are in fact
pulsating stars, although the well-defined double-wave light 
curve of KS~Peg looks very typical for a close binary system.

The absolute magnitudes derived from the HIP parallaxes are listed
in Table~\ref{tab1}. They were determined using:
$M_V = V^{max} + 5 \log \pi - 10$, where the parallax $\pi$ is
in milli-arcsec (mas). They can be compared with the predictions of
the calibration established by \citet{RD97}, 
$M_V(cal) = -4.44 \, \log P + 3.02 \, (B-V) + 0.12$. This
calibration was judged to provide predictions with the mean
standard error
$\sigma (M_V) \simeq 0.22$. The comparison is shown in 
Figure~\ref{fig7} with the $M_V(HIP)$ errors estimated from the
parallax errors, as listed in Table~\ref{tab1} and with
the $M_V(cal)$ errors uniformly assumed at 0.24 as a combination
of the calibration random error and the color index uncertainty.

The $M_V$ calibration is in perfect agreement with the data for
the EW systems, but some deviations are clearly present for EB binaries
and ELL variables. The deviations are expected for the EB systems
as the color index of the dominant companion is not necessarily the
best measure of the combined surface brightness of both components.
In fact, the calibration is surprisingly well behaving even for these
systems. The only case of a large deviation is the system AA~Cet
which has a very poorly determined parallax, obviously due to 
a coupling with the visual-system orbital motion. Since the
system with the observed spectral type and the short period
cannot be as distant as luminous as implied by the 
measured parallax ($M_V(HIP)=0.41$), the value of $M_V(cal)=2.47$ 
has been adopted in this case
for further use. This is the only system in the sample
for which such a substitution was deemed necessary for a proper
evaluation of the luminosity function.

Two ELL systems show large and significant discrepancies between
the two estimates of $M_V$, V335~Peg and IW~Per, in both cases
$M_V(HIP)$ implies a fainter system. We have no explanation for 
these deviations.

\section{THE LUMINOSITY FUNCTION}
\label{LF}

The final sample consists of 32 objects.
Counting in one-magnitude wide $V^{max}$ bins centered on 
$V^{max} = 5, 6, 7$, the numbers are: 2, 6 and 24. 
This progression with the apparent magnitude is very close to the 
expected uniform spatial-density slope
of $4\times$ per magnitude and strongly suggests that our sample is
indeed a complete one. We note that this rate of increase implies
about 100 binaries in the next bin, $7.5 < V^{max} < 8.5$.
This is not observed: the total
number of EW+EB+ELL systems in the combined HIP and 
GCVS catalogs is currently 53. Thus, our choice of limiting the
sample  to $V^{max} \le 7.5$ appears to be a correct one. 

The number statistics of the $M_V$, as listed in Table~\ref{tab1},
can be used to derive the luminosity function by taking 
into account the changing volume for each increment of $M_V$.
%The linear depth changes with the $M_V$; it is
%$d = 25.1$ pc for $M_V = 5.5$, $d = 39.8$ pc for $M_V = 4.5$, etc.
The luminosity function calculated in this way is given in
Table~\ref{tab2} and is shown in Figure~\ref{fig8}.
The errors of the luminosity function have been estimated 
by scaling the Poisson errors by the appropriate
volume. While the faintest bin, $+4.5 < M_V < 5.5$,
contains two systems, the bin $+3.5 < M_V < +4.5$ is empty. 
When plotting the luminosity function in Figure~\ref{fig8}
in logarithmic units,
we arbitrarily entered one system for this bin 
which is consistent with the implied Poisson error for 
such a substitution ($1 \pm 1$ systems); this substitution
was not used in the spatial density estimates below.
Paucity of binaries in the luminosity interval
which is entirely accessible to detection of the
stellar variability strongly illustrates
the strong influence of the Poisson fluctuations for faint systems
of the sample.

The combined spatial density for all EW+EB+ELL systems,
$\rho ({\rm all}) = (1.02 \pm 0.24) \times 10^{-5}$ pc$^{-3}$, is
basically identical to the determination of \citet{due84}. 
The uncertainty of this estimate has been evaluated
by adding the volume-scaled Poissonian error contributions
from all $M_V$ bins. 
The density estimate remains basically the same when very few 
bright systems in the range $-0.5 < M_V < +1.5$ are excluded,
$\rho (> +1.5) = (1.01 \pm 0.31) \times 10^{-5}$ pc$^{-3}$, and 
is reduced only
slightly when one considers only the fainter systems,
$\rho (> +2.5) = (0.90 \pm 0.38) \times 10^{-5}$ pc$^{-3}$.
Thus, the spatial density of contact binaries is dominated
by faint systems, for which the Poisson statistical errors
in the current sample are very large. The current uncertainty
of the spatial density estimate is at the level of a factor 
of about two times. 
 
The Main Sequence luminosity function \citep{wie83}
is shown in Figure~\ref{fig8}, scaled by factor of 500. 
Such a scaled MS function is consistent with the new 
data for luminosity bins with $M_V>+1.5$. 
Table~\ref{tab3} gives the detailed comparison
of the current sample with the scaled MS luminosity function.
The predicted numbers of contact binaries
for the RFO of 1/500 are listed in Table~\ref{tab3} under
$N_{pred}$. The next column in the same table gives the ratio of 
the actually observed numbers, $N_{obs}$ to $N_{pred}$. 
The large Poisson errors of the predicted numbers
indicate that the scaling by the factor of
500 is acceptable for $M_V>+1.5$, with the empty bin of
$+3.5 < M_V < +4.5$ being a 2-sigma fluctuation. 
For the brighter systems within 
$+0.5 < M_V < +1.5$, a further scaling factor of about 10 would 
appear to be applicable, while for $-0.5 < M_V < +0.5$, the additional
scaling factor would be about 60. 

The assumed value for the RFO of 1/500 requires an explanation. 
In fact, the ratio of the luminosity functions, LF(E)/LF(MS),
where E signifies all binaries considered here, gives the RFO
between 1/400 and 1/900 (see Table~\ref{tab3}, the 
empty bin $+3.5 < M_V < +4.5$ is excluded). Only when 
an global average is calculated, one may recover Duerbeck's
value of 1/1000 by inclusion of more luminous objects\footnote{If 
the RFO were really 1/500 for $M_V > +1.5$, the spatial density
of the contact systems would be 
$\rho (1/500) = 1.51 \times 10^{-5}$ pc$^{-3}$, i.e.\
by 50 percent higher than the current estimate, but consistent
with its error.}. The spatial density of contact systems
plunges down for $M_V < +1.5$ (spectral types around A2 -- A4)
to the RFO $\simeq$ 1/5,000, dropping further to some 
1/30,000 for $M_V < +0.5$ (spectral types around B8 -- A0). 
Although the number statistics is poor, the very low RFO 
for short-period binaries at high luminosities is a certainty. 
The very low spatial density of the bright contact systems
was noticed before \citep{ruc98a,ruc98b}, but -- in the
contex of this paper -- is actually expected
given the assumed period limit at one day which naturally 
eliminates all massive and large binaries.

In addition to the luminosity function derived for the current
sample and the scaled MS function, 
Figure~\ref{fig8} shows the luminosity function derived
previously from the 3 kpc deep Baade's Window sample of
Disk Population contact systems using the
OGLE--I data \citep{ruc98b}. We can clearly see a major 
discrepancy in the density estimates by factor of a few times
over the whole range of luminosities. Since the RFO was estimated
before at the level of 1/130 and now we estimate it at about
1/500, the discrepancy involves a large factor of about four
times. We suspect now that the OGLE--I sample either may have 
suffered from an observational problem or that
the discrepancy  tells us something
very important about the difference between the local
population and the one in the Galactic plane,
at distances of 2 to 6 kpc. 
We return to this matter in Section~\ref{discus}.

\section{THE PERIOD FUNCTION}
\label{PF}

The discussion of the orbital period distribution 
for the combined sample of open cluster and Baade's Window 
contact systems \citep{ruc98b} suggested a strong concentration of such 
systems in the period range 0.25 to 0.5 days, with the peak 
around 0.35 days (see Figure~\ref{fig9}).
The peak was found to be even narrower and more concentrated 
at short periods for the binaries in globular clusters \citep{ruc00}.
It may be argued that the short-period concentration 
is an artifact of the use of the linear units 
because -- in this representation and for a flat
$\log P$ distribution (see below) -- the
shorter-period intervals contain relatively more objects. 
However, the linear representation is ideal in demonstrating 
the abruptness of the well established 
cutoff at 0.22 days. We note that this cutoff
still remains entirely unexplained, in spite of several efforts 
\citep{ruc92b,ste95,ste01}.
Most probably, the cutoff is related to the angular-momentum loss
through magnetic activity, but -- paradoxically --  
we do not know which of the two extreme
assumptions is the correct one, i.e.\  whether  
the absence of late-type, short-period contact
systems (Sp$>$K5, $<0.22$ days) results from
the short period binaries being {\it too active\/} to exist for
long time with short periods or {\it not active enough\/} to
appreciably shorten their periods in the Hubble time.

With the current sample, we cannot analyze the period distribution
directly, by simply counting stars,
because of the underlying correlation between the
orbital period and luminosity which requires taking into account
the varying search volume. 
In \citet{ruc98b}, a novel concept of the ``period function'' (PF) 
was introduced. This function is basically the same as the
period distribution except that the number of 
systems is expressed {\it per unit of volume\/}. 
The period function could be
evaluated for the new data by summation of the contributions
in ``layers'' of $M_V$, with the appropriately scaled 
Poissonian errors of the
sums. The results for the current sample are given in Table~\ref{tab2}, 
in both linear and logarithmic units of the period. The functions
are shown graphically in Figures~\ref{fig9} and \ref{fig10}. 
In both figures, the results are compared with those for 
the Baade's Window OGLE--I 3 kpc and 5 kpc sub-samples. We recall
that the 3 kpc sample was better defined for the low-luminosity, 
short-period end of the period distribution, while the 5 kpc 
sample was preferable for $P > 0.6$ days because of the 
improved statistics for
rare high-luminosity objects. It is striking 
how different the OGLE--I and the current samples are. 
Except for the shortest-period bin ($0.2 < P < 0.3$ days) 
and the longest period range ($>0.7$ days),
where the PF's agree within errors, the discrepancy is 
typically at the level of ten times. The missing systems
in the $+3.5 < M_V < +4.5$ luminosity bin may be the main reason for
the depressed PF, but the discrepancy is everywhere large and mainly
reflects the very different estimates of the spatial density from
both surveys. 

Except for the overall scaling relative the OGLE--I
results, the period function derived here is interesting in that it
{\it does not show\/} the short-period peak. 
Bearing in mind the poor statistics at short periods, the function 
is in fact consistent with an assumption of its flatness
in $\log P$ units, which would indicate continuation and extension
of the period distribution as seen at longer orbital periods.
The seminal study of \citet{DM91} showed that the period distribution
of all binary systems, over the very wide range of periods 
can be represented by a Gaussian curve in $\log P$, 
with a shallow maximum at $\log P(d) \simeq 4.8$. As pointed by
\citet{hea96}, the Gaussian part introduces only a small
curvature on distribution which is dominated by the logarithmic
component, $f(P)\,dP = P^{-1}\,dP$. Thus, the flat distribution
$f(\log P)$ would appear to be appropriate and in fact
seems to be valid for the contact systems
discussed in this paper within the relatively short interval 
of $0.25 < P < 1.0$ days. The main point is that we see no
additional ``piling up'' at the very short-period end of this
interval, as was suggested in \citet{ruc98b}.

\section{THE AMPLITUDE DISTRIBUTION}
\label{ampl}

As was shown in \citet{ruc01}, the distribution of the
photometric variability amplitude is a useful tool for studies
of contact binaries. In the present context, it relates 
to the detection constraints for small-amplitude systems; it is
also dependent and carries information on the mass-ratio distribution 
of contact binaries. The examples used in \citet{ruc01} 
included the OGLE--I sample as well as
a preliminary version of the sample discussed in this paper. The
latter was based on objects selected on the basis of the
original Hipparcos Catalog classification and in this 
respect differs from the
current sample which underwent a much more careful scrutiny
for membership and variability type classification.

The amplitude distribution (after correction for presence of third
components in some systems) for the combined EW+EB+ELL 
sample is shown in Figure~\ref{fig11}.
The sample distribution is compared with the calculated
amplitude distribution for the fill out parameter $f=0.25$ and
for a flat mass-ratio distribution, $Q(q) = const$. In contrast 
to the OGLE--I sample, which required a skewed
$Q(q)$, with a strong preference for small $q$,
the current sample does not contradict
the possibility of a flat mass-ratio distribution, although
the statistics is obviously very poor. This discrepancy in the
amplitude distributions can be taken as one of the 
strongest indicators that the OGLE--I sample was 
subject to image blending which systematically
lowered the amplitudes and influenced the distance
determinations.

\section{WHAT WENT WRONG WITH THE DENSITY ESTIMATE FROM THE
BAADE'S WINDOW OGLE--I SAMPLE?}
\label{discus}

The large discrepancy between the current estimate of the
Relative Frequency of Occurrence of about 1/500 (for $M_V > +1.5$)
compared with the RFO derived from the OGLE--I data
\citep{ruc98b}, as was found in Section~\ref{LF}, requires 
an explanation. While the variability-detection
completeness is a strong asset of the current sample, 
the weakness is in its small number statistics. 
As we suggested before, one direct and largest manifestation
of this weakness is absence of any objects in the important
luminosity bin of $+3.5 < M_V < +4.5$, which we suspect 
is a 2-sigma Poisson fluctuation. The current determination of 
the RFO is in an agreement with the estimate of 
\citet{due84} who evaluated the average value
at 1/1000. The results are consistent when allowance is made
for the inclusion of intrinsically bright, but rare, systems in his
sample. The close agreement in the average spatial density
at about $\rho = 1.0 \times 10^{-5}$ pc$^{-3}$
may be fortuitous, but -- most likely --
is a strong confirmation that proper corrections 
have been applied in Duerbeck's determination
for the low-inclination, low-amplitude systems
which were only later discovered by the Hipparcos mission.

Of concern is the large discrepancy of the present estimate
of the RFO with the OGLE--I based result of RFO $\simeq$ 1/130. 
Here we list the several reasons why the Baade's Window sample
has given a different -- and very high -- estimate of the RFO.
\begin{enumerate}
\item For a pencil-beam survey, an error in the distance scale 
enters into a density estimate in the third power. The 
$M_V (\log P, B-V)$ calibration still requires refinements, especially
for extreme colors indices and periods. The Malmquist bias
corresponding to the standard error of the calibration, 
$\sigma (M_V) = 0.22$, is also moderately large; for a uniform
spatial distribution it is expected to lead to a systematic 
error in density of about 30 percent.
\item While the current sample is strictly local, with the depth of
some 40 to 400 pc, depending
on the luminosity range, the OGLE--I survey included all
contact binaries of the Disk Population toward Baade's Window 
within a wide range of distances from about 2 to 8 kpc. 
The nearby systems were excluded 
by the bright limit of the OGLE--I photometry. All statistical 
inferences presented in \citet{ruc97a,ruc97b,ruc98a,ruc98b}
were based on two sub-samples to 3 kpc and 5 kpc 
in depth, but without any nearby systems which were cut off
by the bright limit at around 2 kpc.  
\item Blending of images in the crowded area of Baade's Window
is expected to result in systematic underestimates of distances.
Of particular importance would be underestimation of the distances 
for systems located beyond the respective limits at 3 kpc
and 5 kpc resulting in shifts into the respective sample volumes.  
The blending hypothesis is consistent
with the explanation of the abnormal photometric
amplitude distributions for the two OGLE--I samples, which 
are practically devoid of large-amplitude variables
\citep{ruc01}. Blending of images is very hard to quantify without access
to the original data. We note that even if all binaries were
blended with equally bright stars (a shift in $m-M = 0.75$ mag.),
the increase in the sample depth would be by 1.35 times, whereas
the spatial-density discrepancy demands a factor of $4^{1/3} \simeq 1.59$.
However, simple scaling may be entirely inappropriate as the
numbers of contact binaries were seen to increase dramatically with
the distance so that moderately small errors at the sample far limits 
may have played an enhanced role in density estimates. 
\item The Relative Frequency of Occurrence was utilized in the
OGLE-I data interpretation. While it is a very useful concept,
especially when numbers of variable stars can be directly related
to numbers of stars subject to photometric monitoring 
(as in stellar clusters), the total
number of monitored stars was not easily available
in the particular case of the OGLE-I search, mostly because of the
heavy crowding in this area. Evaluation of the RFO had to be done
only in reference to the predicted density of Disk Population 
stars all the way to the Bulge \citep{ruc98b}, which
in turn depends on the assumed value of the galactic exponential 
length scale, $h_R$, currently a rather poorly known quantity. 
\item While the level of contamination by misclassified
Population~I pulsating stars should be the
same for the current and the OGLE--I samples\footnote{However, the
current sample has an advantage of potential spectroscopic
support in weeding out the pulsating variables.}, the one caused
by the Population~II variables of the RR~Lyr-type and SX~Phe-type 
may be larger in the direction of the Galactic Center. 
One can expect that the number of pulsating stars should be always smaller
than the number of contact binaries (because the former exist only within
the instability strips while the latter can be made from any
MS stars), but this type of contamination cannot be a priori discounted.
The local density of the RR~Lyr stars is $6.2 \times 10^{-9}$
pc$^{-3}$ \citep{sun91}, i.e.\ about 1,600 times lower than the one
derived here for contact binaries. An increase in the spatial density
following $r^{-3}$ or $r^{-4}$ laws would produce an increase by
an order of magnitude or so, hence the RR~Lyr contamination is not 
expected to be very strong.
\item Finally, we cannot exclude a possibility of a genuine 
increase in the number of contact binaries, especially the short-period
ones, in the inner parts of the Galaxy.
The globular cluster sample \citep{ruc00} shows a period
distribution even more strongly skewed to short periods
than the Old Disk OGLE--I sample, with a narrow peak around
0.25 -- 0.5 days. Apparently such systems are not very
common in the solar neighborhood. The inner Disk may contain a 
richer admixture of very old, short-period, low-luminosity systems, 
similar to those found in globular clusters.
\end{enumerate}
None of the above causes is expected to lead to the systematic
error in the space density by a factor of at least 4 times; however,
it is quite possible that several of them actually combined in
elevating the OGLE--I estimate.

\section{FUTURE WORK}
\label{future}

While the present, 7.5 magnitude-limit sample still suffers from
several problems, it is a good starting point for 
establishing a {\it volume-limited\/} sample by extending it deeper
in the magnitude scale. We noted that 
an improvement in the depth by one magnitude, to the limit of
$V^{max} = 8.5$ is expected to result in a substantial increase in 
numbers by about 100 objects. Such objects should be discovered
through searches among bright stars advocated by 
\citet{pac97,pac00,pac01}. We have found that the current 
catalogs contain only about one half of the 
binaries which should be detectable within the one magnitude
interval beyond the current limit, $7.5 < V^{max} < 8.5$. 
It is interesting and instructive to consider what will be these
stars if we indeed increase the limiting magnitude
to $V^{max} = 8.5$. Simply scaling up by factor of four times,
we can predict that most system to be discovered
will be {\it moderately luminous, moderately long-period
(0.4 to 0.7 days), late-A
and early-F type binaries\/} within $+1.5 < M_V < +3.5$.
%We do not expect any major revision in the estimate of the 
%spatial density in this luminosity range.
Some 60 -- 70 binaries of those new 100 systems 
will belong to this group, among them many well-known favorites,
including the prototype, W~UMa.
The brighter and fainter systems will be more difficult to discover: 
At the bright end, for $M_V < +1.5$, 
an immense volume opens up very bright contact binaries, 
but they seem to be genuinely rare in space. We see a strong 
decrease of the spatial density in the present data, which is
partly caused by the adopted period limit of one day 
eliminating massive, large systems, even when on the
Zero Age Main Sequence. Rapid evolution of such systems away
from the ZAMS additionally eliminates compact, early-type
systems. The OGLE--I data suggest a particularly
abrupt cut-off for luminous systems with 
$P > 1.3 - 1.5$ days \citep{ruc98a}. Thus, 
we can expect about 10 -- 15 binaries with $M_V < +1.5$. 
At the other, faint end, with $M_V > +3.5$, probably also few systems
will be detected. Although such systems are
the most abundant in space, the search volume will remain small,
so that deepening of the survey by one magnitude
is expected to yield only some 15 -- 20 intrinsically faint systems.

The implications of these considerations for the best strategy
of establishing the spatial density of contact systems
are as follows: (1) The pencil-beam searches are most promising,
but they must be free of the photometric blending; (2) Wide-sky
surveys are not expected to give immediate results, if conducted
to shallow limiting magnitudes. However, since periods and
luminosities correlate for contact binaries, it may be
possible to conduct targeted (with period constraints)
deeper, wide-field searches. This
may be the best strategy in the sense of a possibility of
applying a combined, photometric and spectroscopic approach to
learn most about these stars through detailed and full analysis.

This paper is expected to provide a departure point for 
further studies. A spectroscopic survey to the same
limiting magnitude would provide a uniform material to derive full
physical parameters of the systems (via $K_i$ radial-velocity
semi-amplitudes). The mass-ratio parameter, $q$, which is
crucial for light-synthesis solutions, would be then 
securely determined as the ratio of the $K_i$ values.
The radial-velocity orbit
data would enable a study of spatial velocities, through
combination of the Hipparcos tangential velocities with the
center-of-mass velocities, $V_0$. Some efforts has been started
in this direction at the David Dunlap Observatory 
where 60 radial velocity orbits, with the accuracy in 
$K_1$, $K_2$ and $V_0$ at the level of about 1~km~s$^{-1}$,
have been recently determined (for 
complete references, see \citet{ddo7}).
This spectroscopic survey very much needs an extension into the
Southern skies. Availability of the mass-ratios will permit
to address the currently missing 
mass-ratio term in the absolute-magnitude 
calibration, $M_V = M_V (\log P, color, q)$ \citep{ruc94,RD97}.

Spectroscopic observations are needed to check the 
variability classification of 14 ellipsoidal systems in the 
present sample. Currently only 6 of them have any spectroscopic data.
There is a possibility that some ELL variables 
may turn out to be pulsating stars.
Gerald Handler (private communication) estimates that about 30 percent of
A/F type Main Sequence stars are $\delta$~Sct stars 
(the same percentage is expected for SX~Phe Population~II stars, 
but these are very rare
in the solar neighborhood). The $\delta$~Sct stars have very short
periods, when compared with contact binaries of the same type
\citep{HS02}, so that they are relatively easy to identify 
with combined spectroscopic and photometric data. However,
variables belonging to the new class of $\gamma$~Dor
\citep{kay99,han99,hen01,HS02}, with pulsation periods from 
about 0.4 days and extending to 3 days,
present a great danger of being taken for ELL binaries. 
Fortunately, most of
them show several pulsation modes simultaneously excited. 

The current situation with spectroscopic data is
summarized in Table~\ref{tab1}. Systems marked SB2 are
double-lined ones which currently have sufficient spectral information.
Systems marked SB1, with one set of lines in the spectra, should be
re-observed because -- for short-period systems -- it is very rare that
modern observing techniques would not show a spectral signature of the
fainter component. All systems marked by a colon should be also
re-observed because the available radial-velocity data are
inadequate.

\section{CONCLUSIONS}
\label{concl}

The current sample of contact (EW) and related (EB and ELL) 
systems with orbital periods shorter than one day and selected from among 
bright stars to $V^{max}=7.5$ (with a large fraction detected
by the Hipparcos satellite) is expected to be complete in the sense
that it contains all binaries showing photometric light variation larger 
than about 0.05 magnitude. This level of photometric accuracy implies
that -- according to predictions of \citet{ruc01} (Section~3.3 
and Figure~4 there) -- only about 10 -- 15 percent of variables
remain undiscovered due to low orbital inclination angle. 
Therefore, even if current statistics
do not take these undetectable systems into account, the
spatial-density estimate of $\rho = (1.0 \pm 0.3) \times 10^{-5}$
pc$^{-3}$ is expected to be sound and consistent with 
the Relative Frequency of Occurrence, RFO $\simeq$ 1/500,
when compared with the Main Sequence stars with $M_V > +1.5$.
While both quantities, $\rho$ and RFO, are still uncertain by a factor 
of about two times, mostly because of the small number statistics,
the estimates should be relatively free of systematic errors.
The spatial density estimate derived here is very close to
that of \citet{due84}, but does not agree with the high value
estimated on the basis of the Disk Population Baade's Window
sample based on the OGLE--I data. As discussed
in Section~\ref{discus}, while the discrepancy  
is most likely due to observational effects which
produced the high value for the OGLE--I sample, we cannot exclude
a possibility that the population of contact binaries observed
towards the Center of the Galaxy is different from the
one in the solar neighborhood.

\acknowledgements

Special thanks are due to Hilmar Duerbeck, Gerald Handler,
Hal McAlister and Bill Hartkopf
for very useful suggestions and comments.

Support from the Natural Sciences and Engineering Council of Canada
is acknowledged with gratitude. 

% \clearpage

\appendix
\section{NOTES ON INDIVIDUAL SYSTEMS}
\label{appendix}

\noindent
The references cited below should be complete as of the end
of the year 2001. 
%Several visually triple systems discussed below were included in 
%\citet[FM2000]{FM00}.

\noindent 
{\bf 44 Boo.} A very well-known, frequently-studied
system, the fainter component of the visual binary
ADS~9494. Another common name i~Boo (preferably
the names should not be combined into 44i~Boo). 
The most general discussion in \citet{hil89a}. 
Many speckle-interferometry studies.  
The newest, more accurate radial velocity orbit in \citet{ddo4}.
The data on third-body light contribution from 
\citet{FM00}. The spectral type for the binary is 
estimated at K2V from the color index $B-V=0.94$. The combined 
spectrum is dominated by the brighter G1V companion.

\noindent   
{\bf VW Cep.} A well-known, frequently-studied
system. The most extensive study with references to
earlier studies, \citet{HM00}.
The data on third-body light contribution from \citet{FM00}. 

\noindent
{\bf V759 Cen.} Fainter at light maximum than 7.5 mag.\ 
limit in $H_P$, but slightly brighter in $V$, in accordance
with the discovery data of \citet{bon70}, $V^{max} \simeq 7.45$.
Lacks a radial velocity study. The HIP light curve poor, possibly
light curve changes. 

\noindent
{\bf HT Vir.} In a tight (0.56 arcsec) visual triple system,
the slightly brighter component of ADS~9019. 
Excluded from the sample since an allowance for the 
third star has pushed its 
$V^{max}$ to 7.86. A very interesting system, with a good
radial-velocity orbit in \citet{ddo4}.

\noindent
{\bf KR Com.} Discovered by Hipparcos. No reason for the type EB
as in the HIP catalog. A radial velocity study in \citet{ddo6}. 
Many speckle-interferometry studies of the visual binary
ADS~8863. Excluded after allowance for the light of the third star.

\noindent
{\bf V566 Oph.} A well-known, frequently-studied system. 
An extensive spectroscopic study with earlier references
in \citet{hil89b}.

\noindent
{\bf VW Pic.} Discovered by Hipparcos. A triple system with the fainter 
(1.9 mag.) visual companion separated by 9.9 arcsec.
Called EB in HIP, but the
eclipses are of equal depth. The Fourier decomposition of the light
curve gives the type EA, so the this system has been
excluded from the sample. 

\noindent
{\bf AW UMa.} One of the most frequently studied W~UMa-type systems,
mostly because of its extreme mass-ratio of $q=0.08$ and
presence of total eclipses. Recently
relegated to the second place in the extremity of the mass-ratio,
in favor of SX~Crv \citep{ddo5} with $q=0.066$. 
The most recent spectroscopic data in \citet{ruc92a}.

\noindent
{\bf V972 Her.} Discovered by Hipparcos. Called EB in HIP, but 
the light curve too shallow for classification, so classified as ELL.
The spectroscopic orbit \citep{ddo6} indicates a low inclination systems.

\noindent
{\bf V445 Cep.} Discovered by Hipparcos, but the period is in fact
two times longer than given in HIP.

\noindent
{\bf CN Hyi.} Discovered by Hipparcos. Included in \citet{FM00}, but only
one visual component given, so most probably not a visual triple.
A nice, well defined HIP light curve.

\noindent
{\bf MW Vir.} Discovered by Hipparcos, but the period in fact
two times longer than given in HIP.

\noindent
{\bf V867 Ara.} Discovered by Hipparcos, but the HIP light curve
rather poorly covered. A member of a visual binary 
\citep{FM00}. Excluded from the sample because
the Fourier decomposition of the light
curve suggests the variability type EA. 

\noindent
{\bf KS Peg.} Called EB in HIP, but the light variation too small
to classify the type. \citet{HG85} and \citet{hub88} presented spectroscopic
and photometric observations and suggested an extreme mass ratio,
$q < 0.1$. The currently ongoing DDO radial velocity observations 
encounter difficulties as the broadening function is very
wide, but does not split. The system requires an in-depth analysis.

\noindent
{\bf AA Cet.} Discovered by \citet{blo71} who classified the binary as
EW and found the secondary eclipse total. The brighter component of
a visual binary ADS~1581 with the separation of 8.4 arcsec
\citep{tok97}. The (first published)
light curve obtained by Hipparcos is definitely of an EB binary. This
bright binary requires both photometric and spectroscopic study. The
secondary eclipse may be total.

\noindent
{\bf V918 Her.} Discovered by Hipparcos. No other data currently 
available. Called EB in HIP, but included among ELL here.

\noindent
{\bf V353 Peg.} Discovered by Hipparcos. No other data currently 
available. Called EB in HIP, but included among ELL here. 

\noindent
{\bf $\epsilon$ CrA.} The brightest $V^{max}=4.73$ and one of the most 
frequently observed contact systems. 
% Discovered by \citet{CC50}.
The HIP light curve is rather sparsely
covered. Light curve studies of \citet{tap69}, \citet{her72},
\citet{twi79} gave $q_{phot}=0.113 \pm 0.002$, while the spectroscopic
study of \citet{GD93} gave $q_{spec}=0.128 \pm 0.014$, so no discrepancy
in the mass ratio here, confirming good quality of data. 

\noindent
{\bf RR Cen.} 
%Discovered in 19th century by \citet{rob1896} and 
%frequently studied since. 
Major photometric studies: \citet{kni61}, \citet{kni65},
\citet{cha71}, \citet{MD72}, \citet{twi79} 
gave consistent results, mostly because of the presence of total eclipses. 
The most recent spectroscopic study by \citet{KH84} could be improved
because of the low spectral resolution used.

\noindent
{\bf IS CMa.} A large amplitude W~UMa-type system discovered by HIP.
The system begs further work.

\noindent
{\bf V851 Ara.} Discovered by HIP. A shallow and sparse HIP light curve.
No other data available except being included in the search for $\lambda$ Boo
stars \citep{pau01} with negative result. Called EB in HIP, 
but the light variation too small to classify the type, so called ELL. 

\noindent
{\bf LL Vel.} 
%Discovered by \citet{lam87}. 
The variability type uncertain \citep{mat90,KH95,MP95} 
because of the very small light variations, so
ELL as in HIP is confirmed.

\noindent
{\bf V535 Ara.} 
%Discovered by \citet{str64}.
Photometric studies by
\citet{cha67} and \citet{sch79}. An UV light curve by \citet{eat91}.
The spectroscopic orbit of \citet{sch79} certainly leaves a lot
of room for improvement.

\noindent
{\bf S Ant.} Not much work has been done on this nice EW system since
the classic photometric and spectroscopic study of \citet{pop56}
where the binary was described as a single-lined system. It should be
easy to determine a nice SB2 spectroscopic orbit with a
modern equipment.

\noindent
{\bf V1010 Oph.} 
%Discovered by \citet{str64e}. 
Photometric studies
by \citet{leu74} and \citet{LW77}. Three SB1 spectroscopic orbits 
summarized in \citet{wor88}. An attempt to detect spectral lines of
the secondary should be made. 

\noindent
{\bf V449 Aur.} Discovered by Hipparcos. Excluded as the HIP light curve
suggests a detached binary with shallow eclipses.

\noindent
{\bf 2082 Cyg.} Discovered by Hipparcos. No other data currently 
available. 

\noindent
{\bf KP Peg.} Discovered by \citet{wal88} as a variable member of
the visual binary ADS~14977 \citep{tok97}. Light curve in \citet{kes89}.
% [check AApS,96,137,1992; ApSS,155,113,1989]

\noindent
{\bf PP Hya.} Discovered by Hipparcos. No other data currently 
available. 

\noindent
{\bf FO Vir.} Discovered by \citet{egg83} and extensively observed
in the following years \citep{SF84,ant84,por87}. \citet{moc86}
estimated photometric mass ratio at 0.15, but the secondary has not
been discovered spectroscopically yet.
The spectral type of \citet{GG89} indicates that the star is
slightly metal-poor. The star appears as CCDM~J13298+0106A among
visual binaries in the HIP database, but no other information 
is currently available.

\noindent
{\bf V1130 Tau.} Discovered by Hipparcos. The light curve looks like that
of two strongly distorted, identical stars, but not of a contact binary, and
this is confirmed by the Fourier decomposition of the light curve. The
recent DDO spectroscopic data (Rucinski et al., private communication)
indicate that it is a detached binary and definitely not an EB system.
It is a weak-metal star \citep{abt86,GG89}. 

\noindent
{\bf V2388 Oph.} \citet{rod98} discovered variability of the star, 
apparently independently of the HIP discovery. The star was a 
subject of many speckle interferometry investigations before 
discovery of the photometric variability. The revised visual binary orbit
given by \citet{soed99}. The spectroscopic study of \citet{ddo6}
confirms the relative luminosity estimates of \citet{soed99} and
gives a good radial-velocity solution for both components.

\noindent
{\bf V335 Peg.} Discovered by Hipparcos. No other data currently 
available. The photometric variability very small so that the photometric
period is uncertain and may require revision.

\noindent
{\bf TV Pic.} 
% Discovered by \citet{ver87}. 
The light curve shows a large
O'Connell effect of the first maximum being higher then the 
second maximum. The photometric and high-resolution 
spectroscopic study of \citet{pav98} discusses peculiarity of the
system, but no radial-velocity orbit yet available. The binary is
similar to TU~Hor.

\noindent
{\bf ES Lib.} 
%Discovered as a variable star by \citet{str64x}. 
A photometric study and a single-line radial-velocity orbit at low
spectral resolution by \citet{bar73}. The star has been a subject of 
several speckle interferometry studies. H.\ A.\ McAlister 
(private comm.) estimates
that the visual companion must be faint, with $\Delta m > 2.5$, which
is confirmed by the apparent large (``undiluted'')
photometric amplitude of ES~Lib.

\noindent
{\bf IW Per.} 
%[discovery? Kim 1980 ApSS,68,355]
Long recognized as an ellipsoidal variable \citep{mor85}. 
Recent interest concentrated on the chemical peculiarity
\citet{ade98,PM98}. \citet{AM95} list among the Am stars
and estimate $V \sin i = 91$ km~s$^{-1}$.

\noindent
{\bf DW Boo.} Discovered by Hipparcos. No other data currently 
available. Classified as EB in HIP, but this classification should
be replaced by ELL in view of the very small light variations.

\noindent
{\bf TU Hor.} 
%[who discovered?]. 
In the list of ELL variables
of \citet{mor85}. The light curve shows a large O'Connell effect 
which very strongly contributes to the combined light variations. 
A photometric study by \citet{due77}. The spectroscopic
single-lined orbit by \citet{due79} [check] 
implies a low orbital inclination.

\noindent
{\bf GK Cep.} Much studied system previously considered EW or EB, but
excluded by us because the Fourier light curve decomposition indicates
a detached binary. Pervious references in \citet{nia91}.

\noindent
{\bf HL Dra.} Discovered by Hipparcos. No other data currently 
available. The Fourier light curve decomposition indicates
a detached binary, thus not confirming the EB type given in HIP.

\noindent
{\bf DX Aqr.} The Fourier decomposition of the rather poorly
covered HIP light curve indicates a detached binary. Member of the
visual binary ADS~15562 \citet{tok97}.
A radial velocity study by \citet{PS76}.

\clearpage

\noindent
Captions to figures:

\bigskip

\figcaption[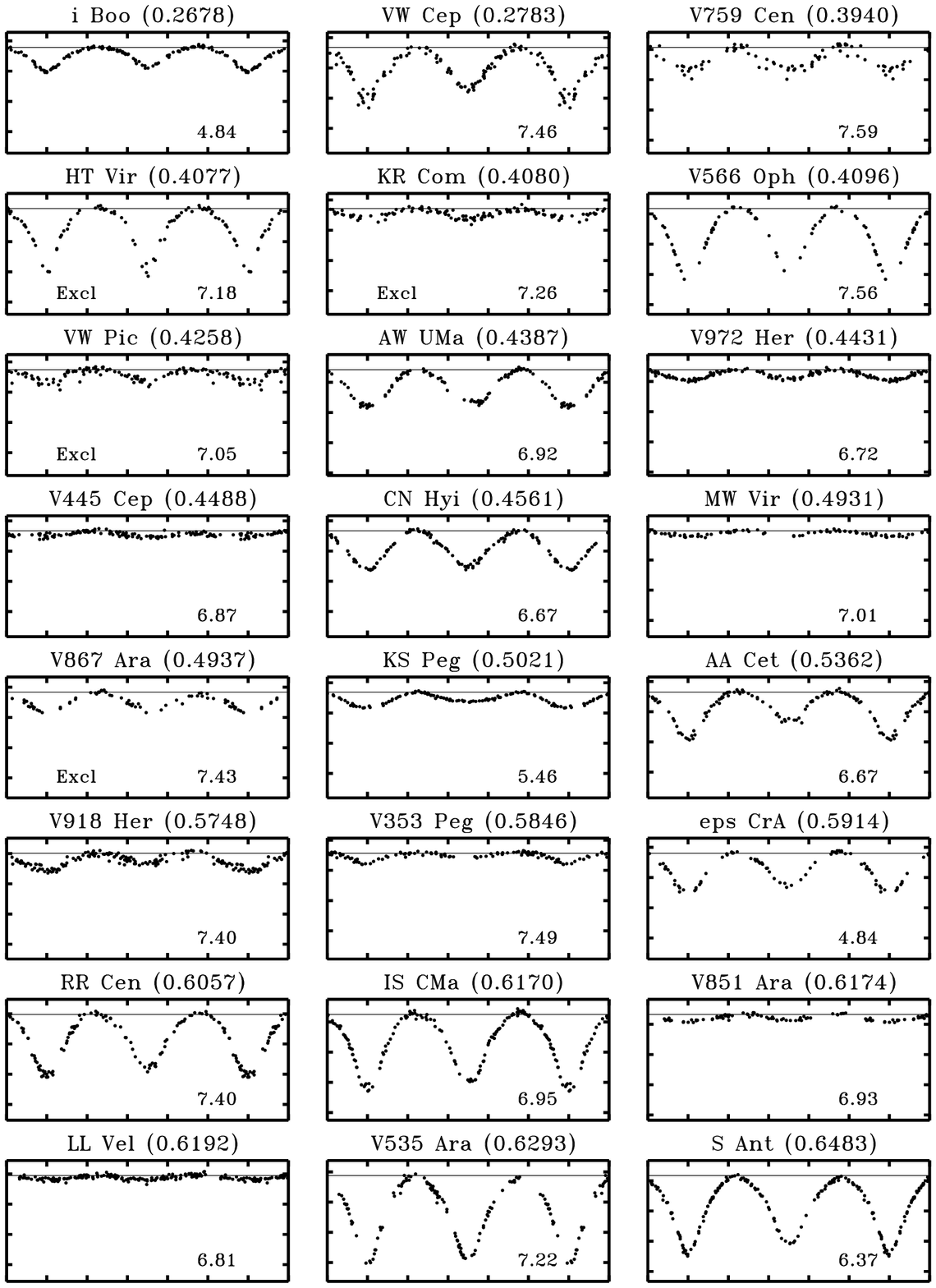] {\label{fig1}
Hipparcos light curves of eclipsing systems considered 
for the final sample ($H_P$ magnitudes). 
The orbital periods in days are given in
parentheses by the names of the systems.
The thin horizontal lines show the adopted maximum-light
magnitudes $H_P^{max}$, which are given in numbers within the boxes. 
For those systems which did not have previously
measured maximum-brightness $V$ magnitude,
these have been adopted for transformations
into $V^{Max}$; two systems fainter in $H_P$ than
7.5 mag., V759~Oph and V566~Oph, have been retained
in that process. Three systems, HT~Vir, KR~Com and VW~Pic
have been excluded because they were fainter than $V^{max} = 7.5$
when allowance was made for their companions. V867~Ara was found
to be a detached binary, although its light curve is poorly
covered and the EA classification may have to be revised.
}

\figcaption[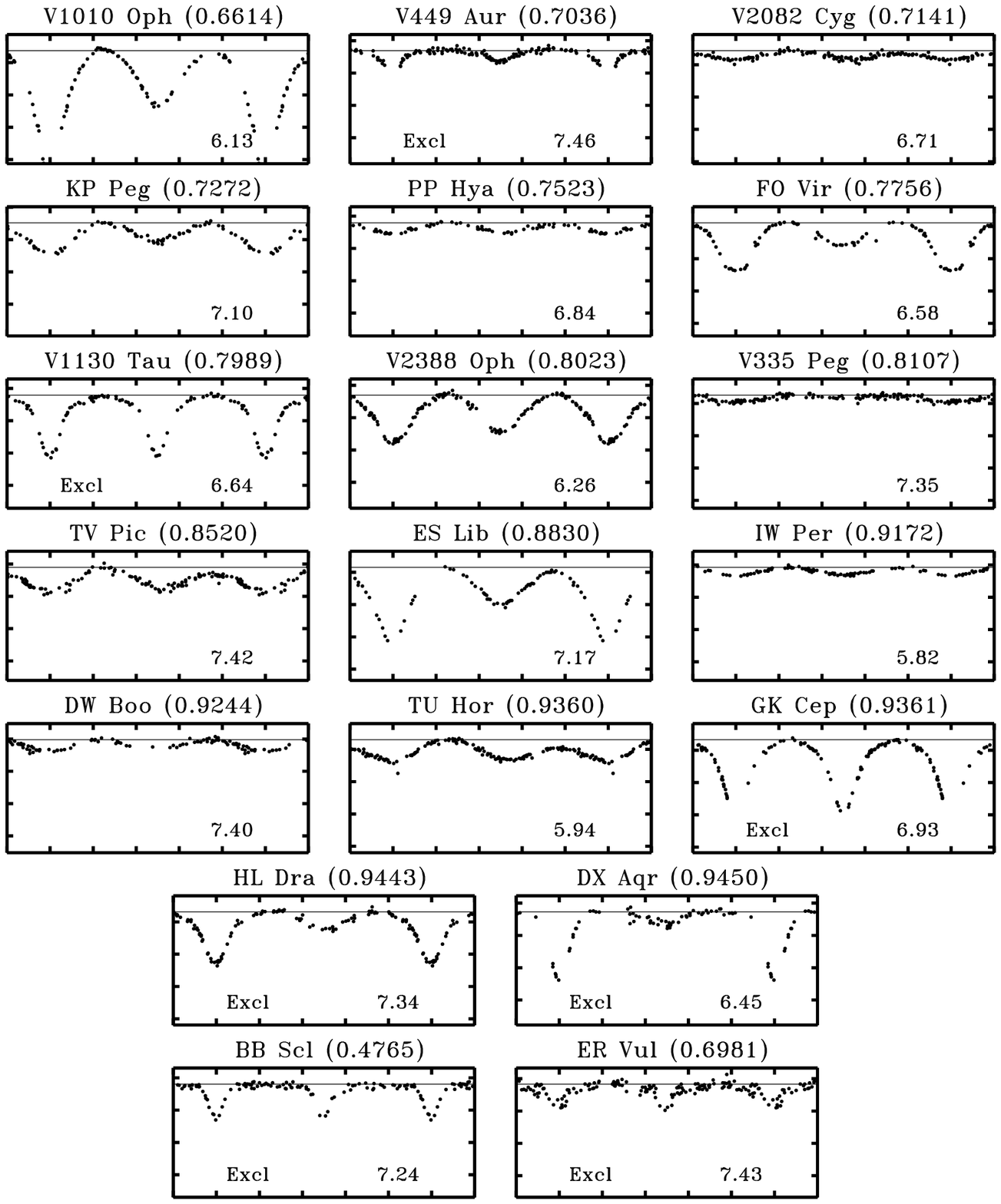] {\label{fig2}
The same as in Figure~1 for the remaining systems.
V449~Aur, V1130~Tau, GK~Cep, HL~Dra and DX~Aqr
were found to be a detached binaries. BB~Scl and ER~Vul, shown
at the bottom of the figure, are the only two detached
binaries fulfilling the period and brightness criteria with
the available Hipparcos light curves. They are shown here 
for illustration purposes only.
}

\figcaption[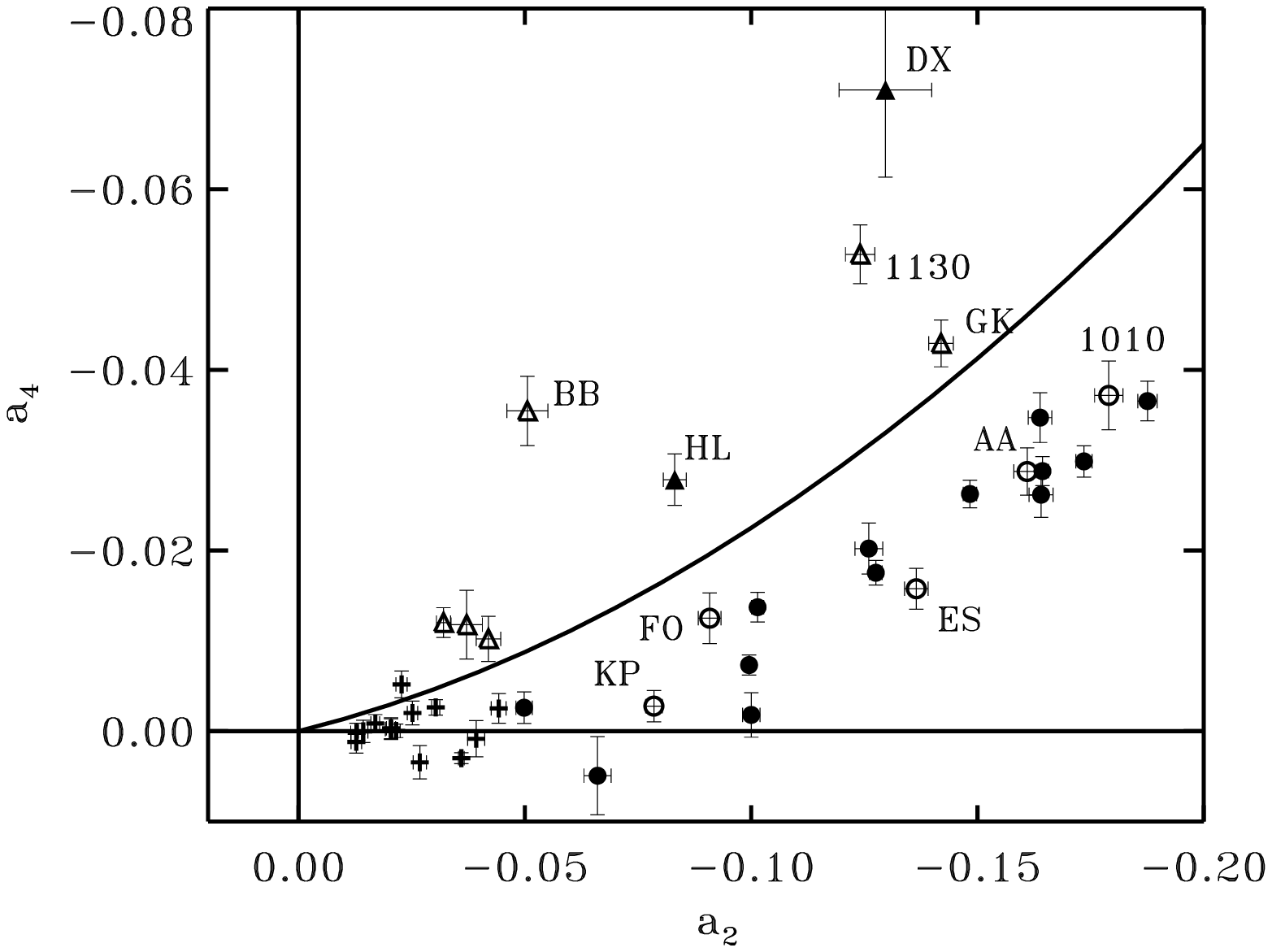] {\label{fig3}
The Fourier coefficient $a_2 - a_4$ have been used to
separate detached binaries (above the curve, triangles) from the EW 
(filled circles) and EB (open circles) binaries. The errors
of the Fourier cosine coefficients have been determined by the
bootstrap sampling process. In some cases, when the system
happened to have a visual companion, the coefficients and their
errors became magnified after allowance was made for additional
light in the system. The open (EB) and closed (EW) symbols relate to
the relative depth of eclipses, as defined in the next Figure~4.
Some systems are labeled by their variable-star names. 
}

\figcaption[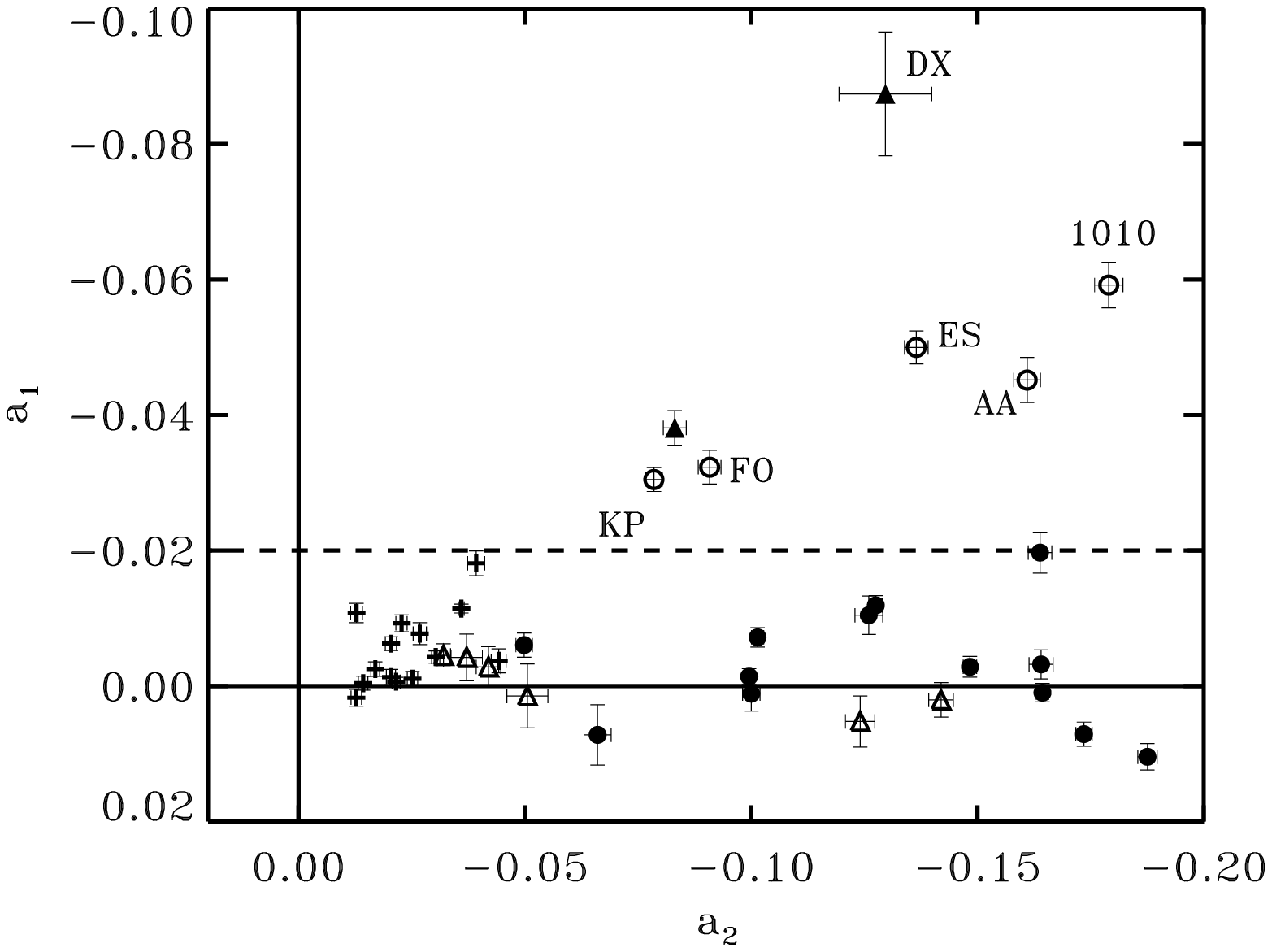] {\label{fig4}
The Fourier coefficients $a_2 - a_1$ used to
separate EB binaries (open circles) from the rest of the
systems. The other symbols are as in Figure~3.
}

\figcaption[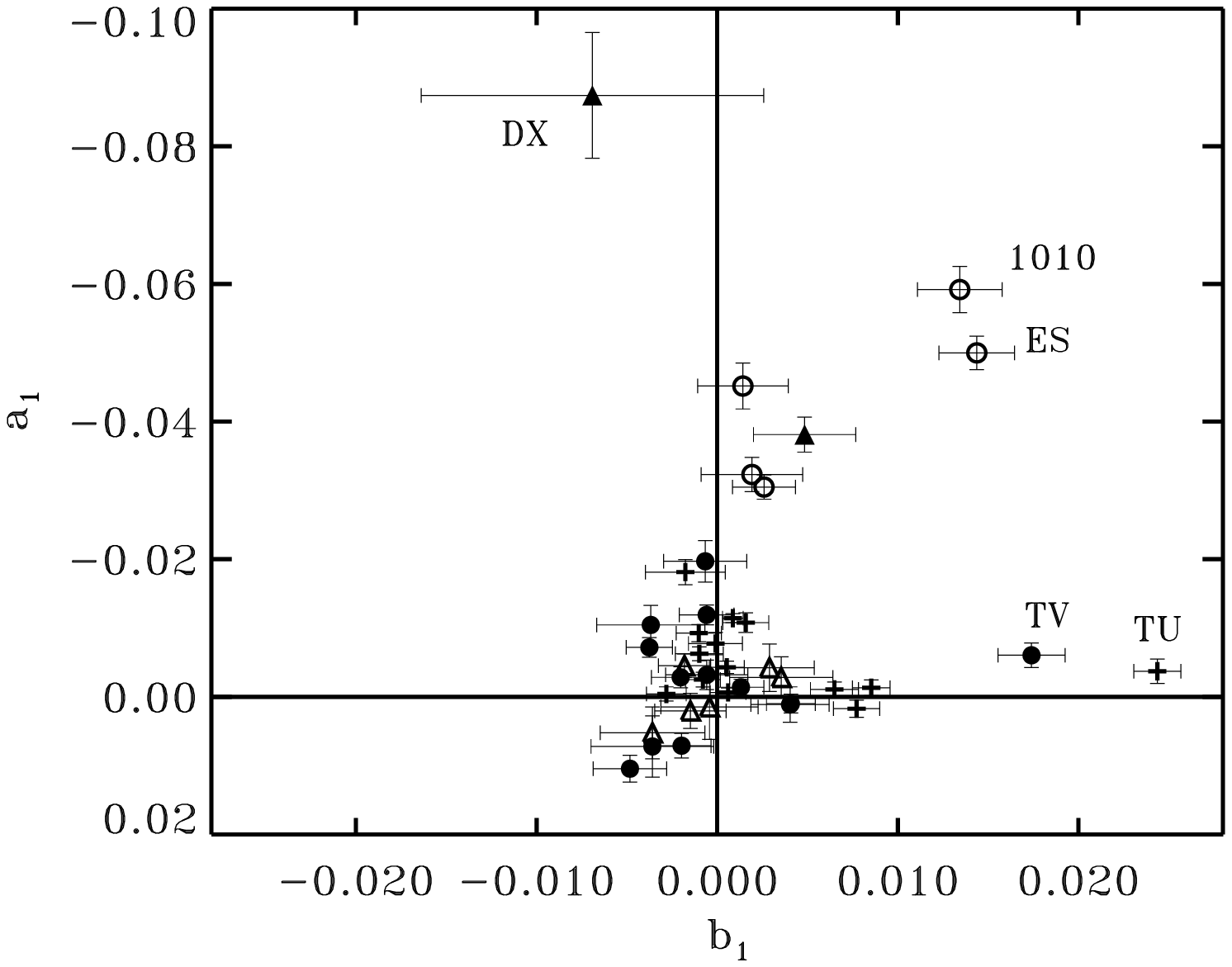] {\label{fig5}
The Fourier coefficients $b_1 - a_1$ used to
find systems with light-curve maximum asymmetry (as measured
by the first sine coefficient $b_1$). 
The symbols used are the same as in the previous two
figures. The observed correlation for ES~Lib and V1010~Oph 
suggests that these are semi-detached binaries transferring
matter from the more- to the less-massive components, with
an accretion spot on the side of smaller component.
}

\figcaption[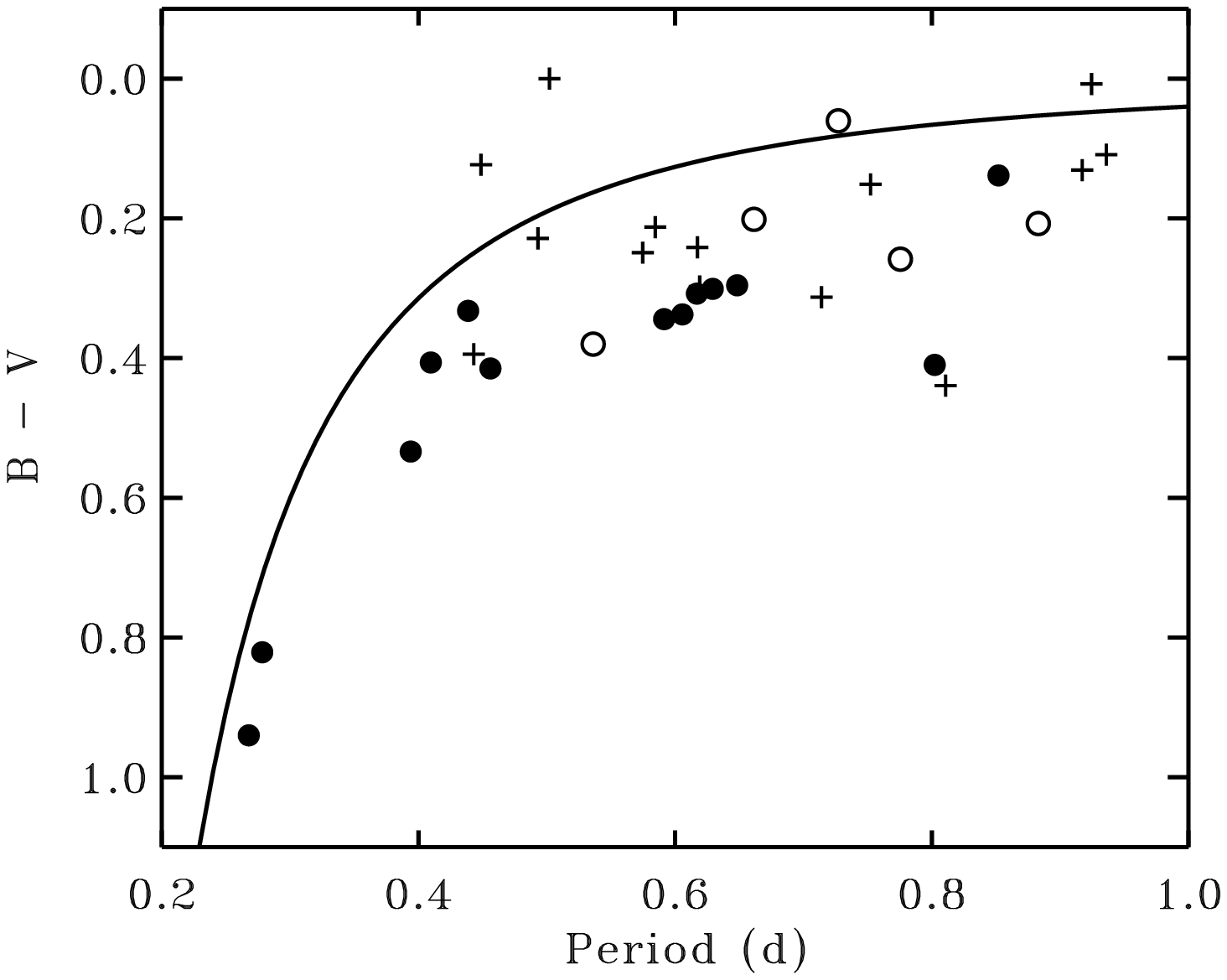] {\label{fig6}
The period--color relation for the current sample.
The symbols for the EW, EB and ELL systems
are the same as in the previous figures. The continuous line
gives the Short-Period Blue Envelope (SPBE) as derived in
\citet{ruc98b}; evolution and reddening are expected to
move stars down and to the right, away from the curve.
Two ELL systems above the SPBE are V445~Cep and KS~Peg.
}

\figcaption[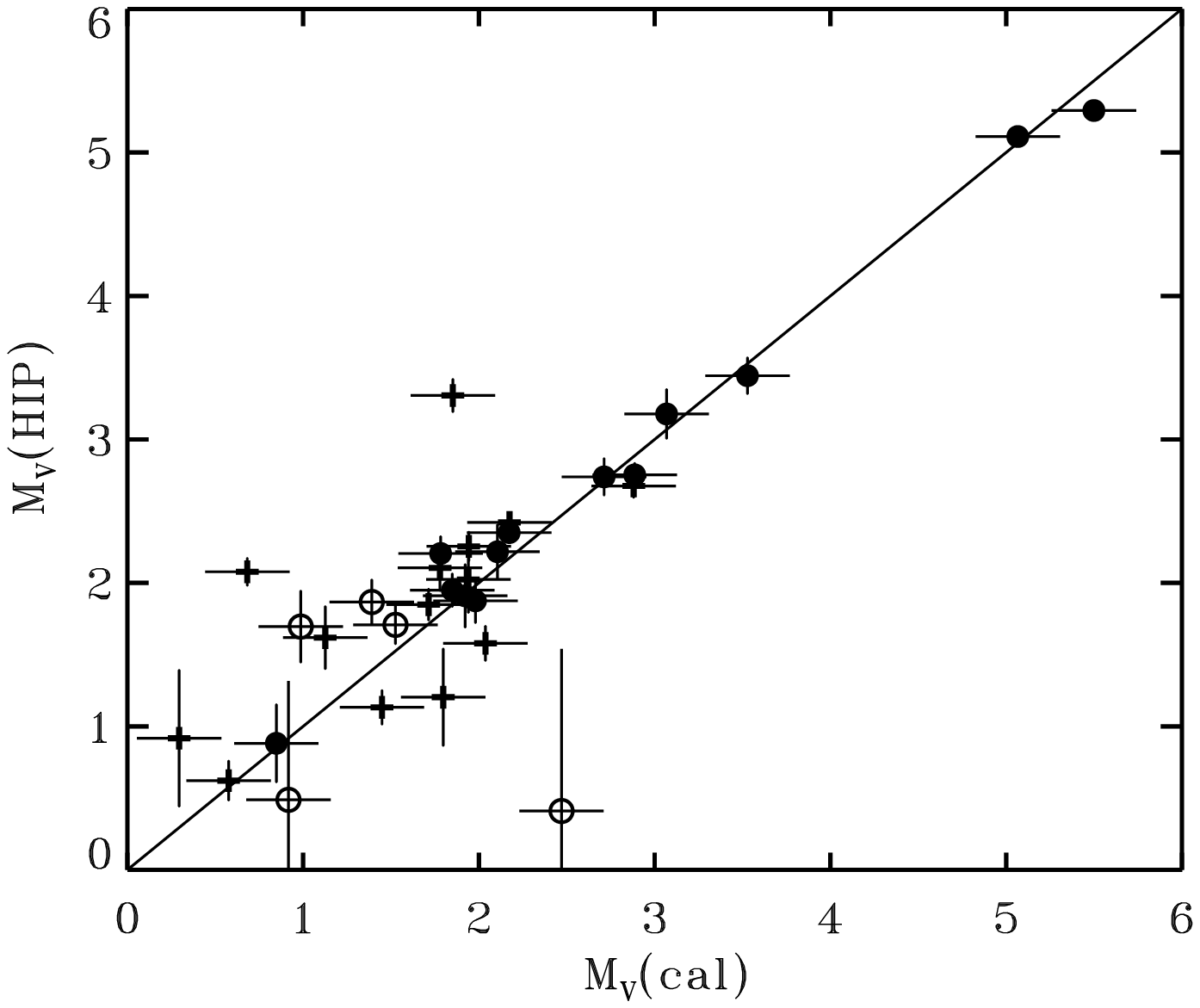] {\label{fig7}
Comparison of the absolute magnitudes determined from the
Hipparcos parallaxes with those estimated from the
\citet{RD97} calibration. The symbols are the same as
in the previous figures. The error bars for $M_V(HIP)$ are
from the errors of the parallaxes while the errors in
$M_V(cal)$ are assumed the same at 0.24 mag. 
}

\figcaption[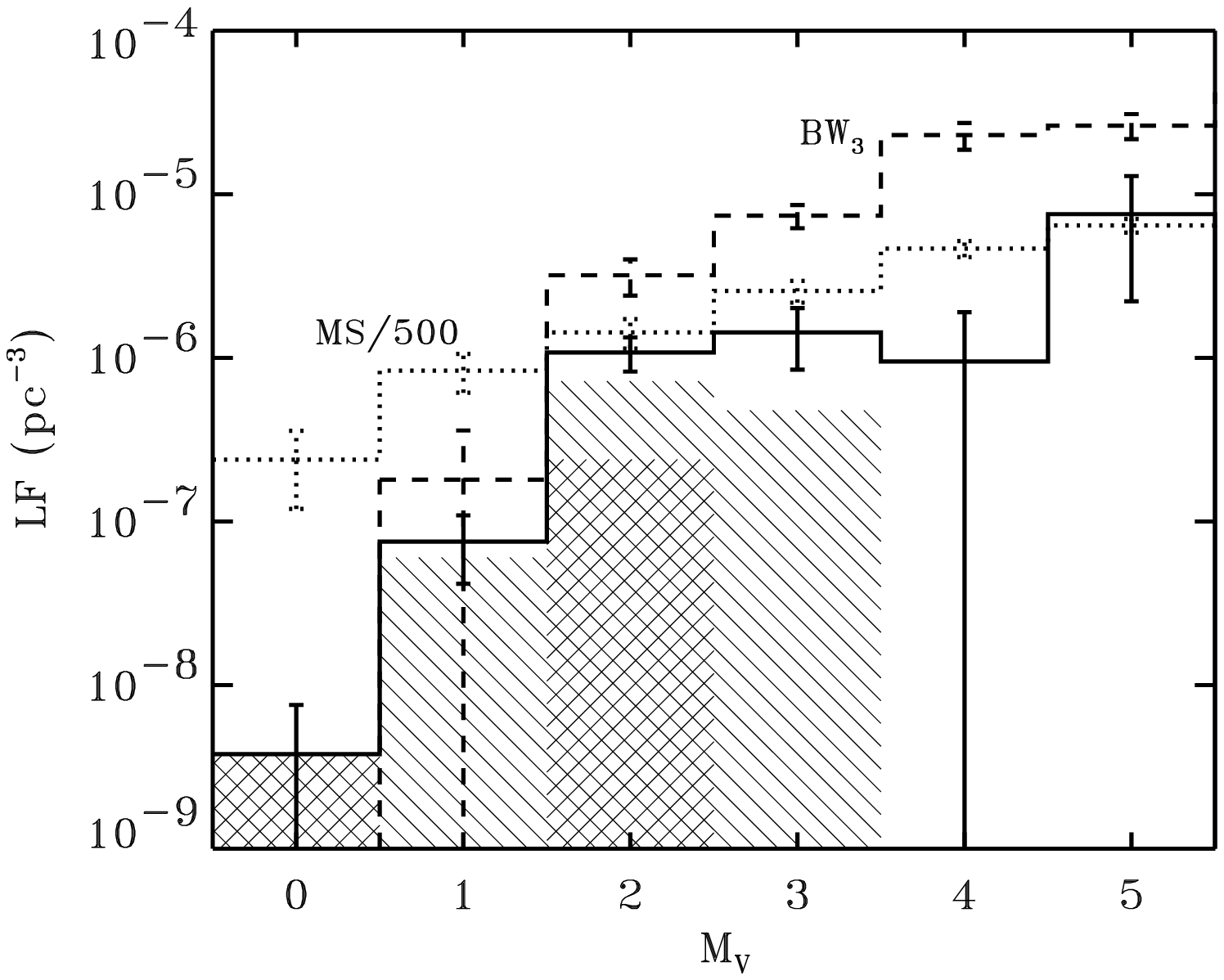] {\label{fig8}
The luminosity function for the whole sample 
(thick line) and for the sub-samples of ELL and 
EB systems (shaded and doubly shaded areas, respectively).
The error bars show the Poisson errors in the bins.
The Main Sequence function, scaled by factor 500, is shown by
a dotted line. The OGLE--I estimate for the 3 kpc
Baade's Window sample \citep{ruc98b} is shown by
the broken line. The error bars are based on Poisson statistics.
Note that absence of low-luminosity EB and ELL systems
suggested by this plot is not significant and may be explained
by the low number statistics for faint systems. 
}

\figcaption[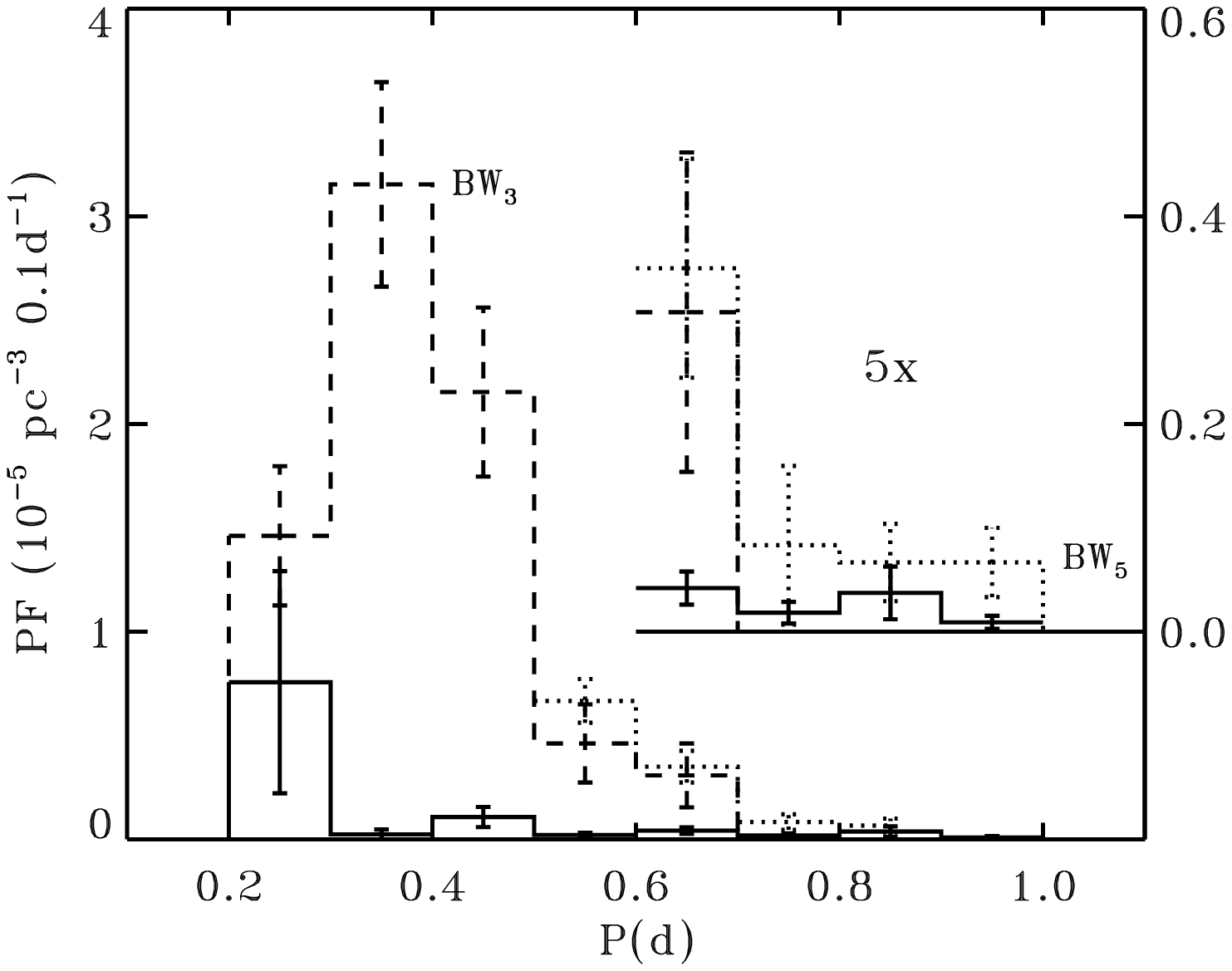] {\label{fig9}
The period function for the combined sample (thick line) 
is compared with the data derived from the 
OGLE--I results \citep{ruc98b}. The broken line shows
the period function for the 3 kpc sample ($BW_3$), 
while the dotted line shows the same for the 
larger 5 kpc sample ($BW_5$). The data
for the latter are valid only for $P>0.5$ days,
but are statistically better defined than for the 3 kpc 
sample in this period range.
The insert shows the long-period portion magnified by factor 
$5\times$ (right vertical axis) to illustrate that, for long periods,
the current period function is basically consistent with that
for the 5 kpc OGLE-I sample. 
Note that linear units of the period are used in this plot. The
next figure shows the same data in the logarithmic units.
}

\figcaption[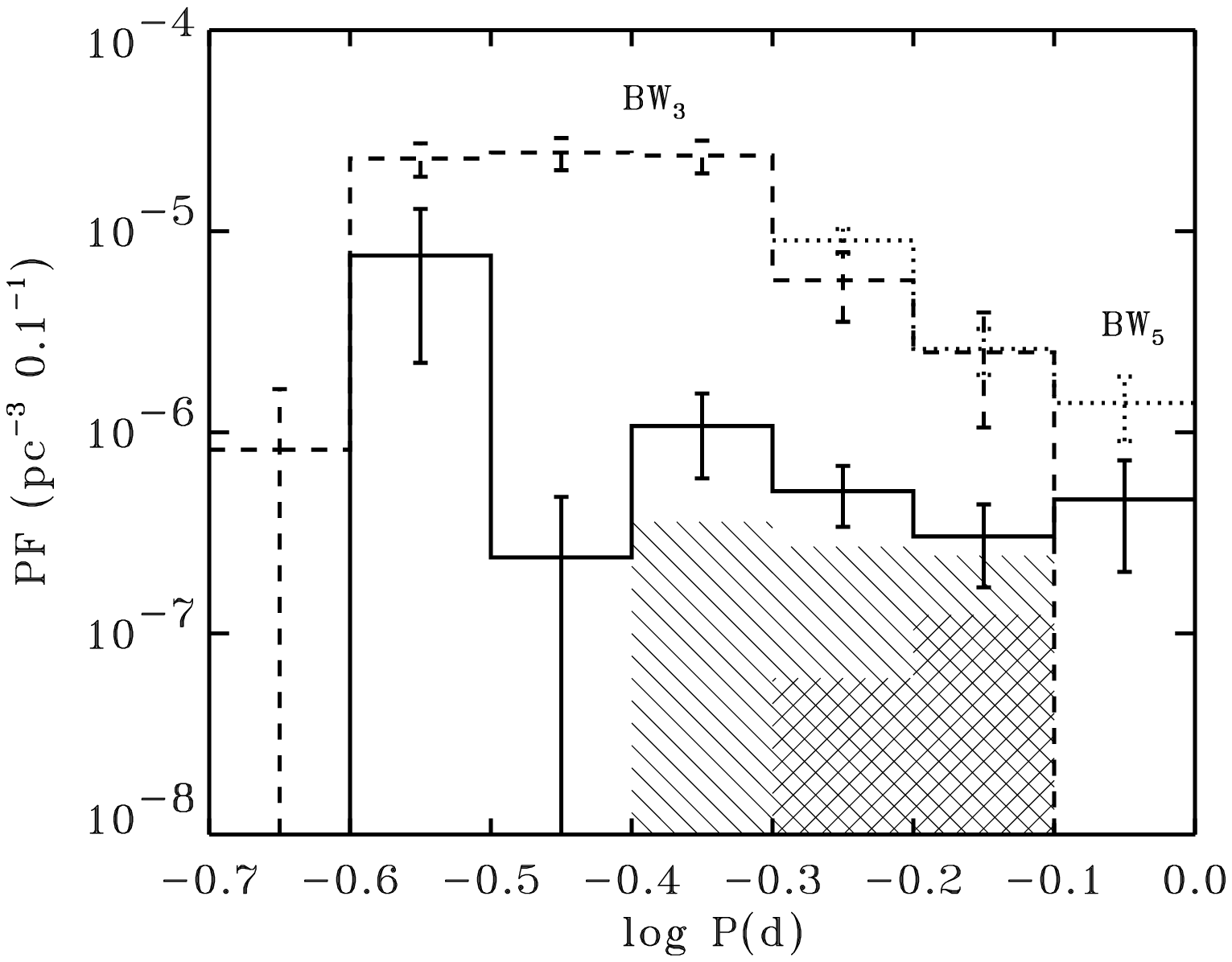] {\label{fig10}
The period function for the whole sample (thick line), 
in the logarithmic units of the period. 
The sub-samples of ELL and EB systems are marked
by shaded and doubly shaded histograms, as in Figure~\ref{fig8}
(see the caption to this figure).
The OGLE--I 3 kpc ($BW_3$) and 5 kpc ($BW_5$) samples 
\citep{ruc98b} are shown by the broken and dotted lines. 
See the text and the previous figure for further explanations.
}

\figcaption[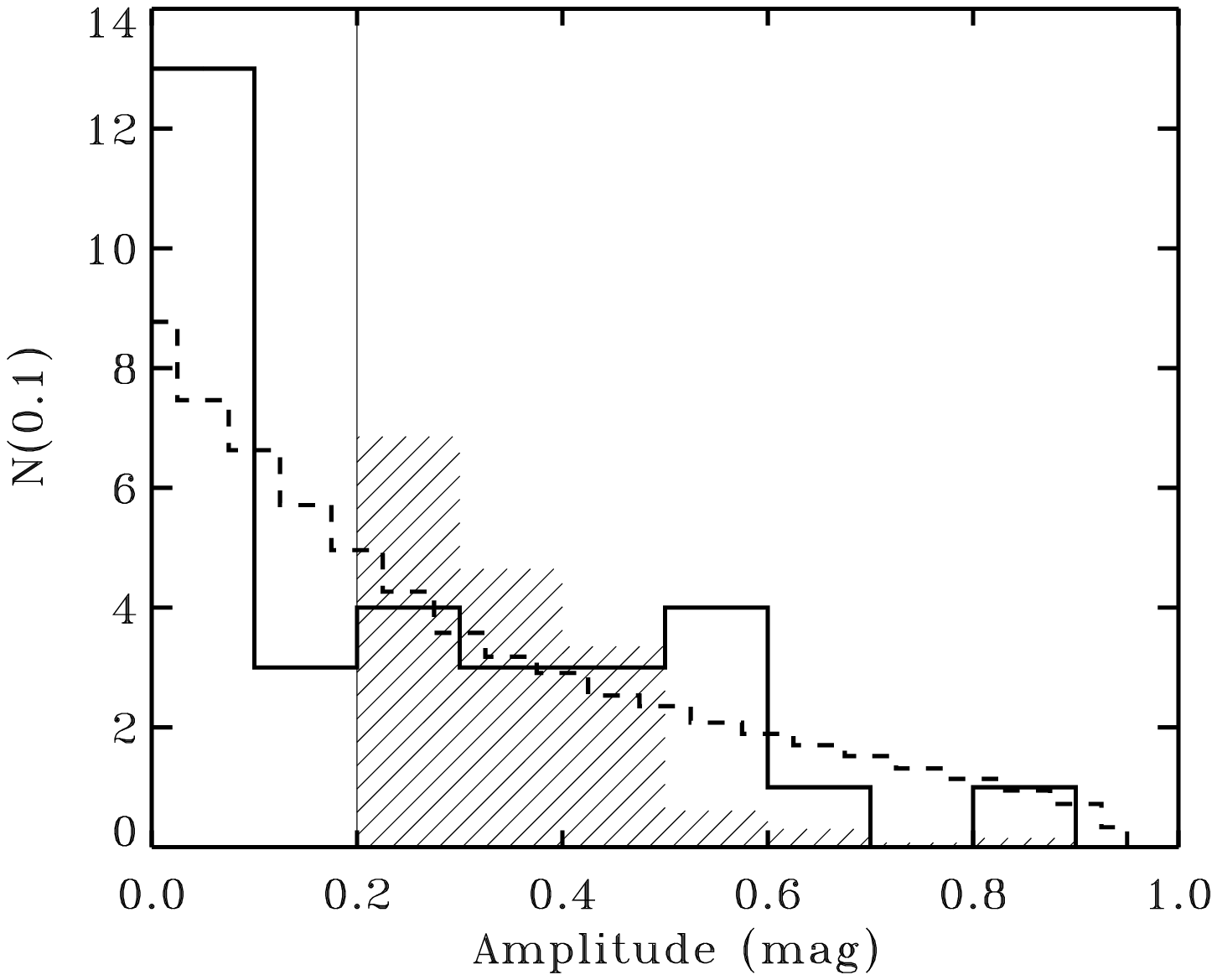] {\label{fig11}
The amplitude distribution for the sample. For systems in triple
systems, the corrected amplitudes are shown. The ELL binaries
occupy, basically by definition, the first bin at $<0.1$ mag.\
because the criterion for ELL assignment 
was on the Fourier term, $-0.05 < a_2 < 0$.
%The only exception is the system TU~Hor with a strongly asymmetric light
%curve with the total amplitude is 0.13 mag.
%The shaded bars give the contributions of ELL and EB binaries.
The dotted line gives the calculated distribution 
for contact binary systems \citep{ruc01} with the degree of
contact $f=0.25$ and the flat mass-ratio distribution, normalized to
the same total number of objects.
The shaded area gives the distribution for the OGLE Baade's
Window sample to 5 kpc (which includes the 3 kpc sample), 
truncated to and normalized for 
objects with amplitudes larger than 0.2 mag.
}

\clearpage                    % must be here for labels to be recognized

\begin{table}                 % large table of individual systems
\dummytable \label{tab1}      % Table 1
\end{table}

\begin{table}                 % table with LF & PF
\dummytable \label{tab2}      % Table 2
\end{table}

\begin{table}                 % table with RFO
\dummytable \label{tab3}      % Table 3
\end{table}

\end{document}

\setlength{\textwidth}{10.0in}
\setlength{\textheight}{6.8in}
\setlength{\hoffset}{-0.6in}
\setlength{\voffset}{-1.6in}

\begin{deluxetable}{lrrcccccccccccccccccll}

%\tabletypesize{\small}
\tabletypesize{\scriptsize}

%\rotate
\pagestyle{empty}
\tablecolumns{22}

%\tablewidth{900pt}
\tablewidth{0pt}

\tablenum{1}
\tablecaption{The 7.5 magnitude sample of EW+EB+ELL systems}
\tablehead{
   \colhead{Name} & \colhead{HIP} & \colhead{HD} & \colhead{P} &  
   \colhead{H?} & \colhead{EX?} & \colhead{Type} & 
   \multicolumn{2}{c}{$V^{max}$} &
   \colhead{$B-V$} & \colhead{$\beta$} & \colhead{sep} &
   \colhead{ampl} &  \colhead{$\pi$} & \colhead{$M_V$} &
   \colhead{$a_0$} & \colhead{$a_1$} & \colhead{$a_2$} &
   \colhead{$a_4$} & \colhead{$b_1$} & \colhead{RV} & \colhead{Sp} 
\\
   \colhead{(1)} &  \colhead{(2)} &  \colhead{(3)} &  \colhead{(4)} &
   \colhead{(5)} &  \colhead{(6)} &  \colhead{(7)} &  \colhead{(8)} &
   \colhead{(9)} &  \colhead{(10)} &  \colhead{(11)} & \colhead{(12)} &
   \colhead{(13)} &  \colhead{(14)} &  \colhead{(15)} &  \colhead{(16)} &
   \colhead{(17)} &  \colhead{(18)} &  \colhead{(19)} &  \colhead{(20)} &
   \colhead{(21)} & \colhead{(22)}
}
\startdata
i Boo    &  73695 & 133640 & 0.26782 &   &    & EW & 4.73 & 5.82 & 0.55 & 1.74 & 1.70 & 0.17 & 78.4 $\pm$ 1.0    & 5.29 & 0.953 & $-0.007$ & $-0.060$ & $-0.013$ & $ 0.000$ & SB2 & K2V \\
VW Cep   & 101750 & 197433 & 0.27831 &   &    & EW & 7.32 & 7.32 & 0.82 & 0.06 & 0.50 & 0.35 & 36.2 $\pm$ 1.0    & 5.11 & 0.892 & $-0.010$ & $-0.118$ & $-0.019$ & $-0.003$ & SB2 & K0V \\ 
V759 Cen &  69256 & 123732 & 0.39400 &   &    & EW & 7.47 &      & 0.53 &      &      & 0.15 & 15.9 $\pm$ 0.9    & 3.44 & 0.934 & $+0.007$ & $-0.066$ & $+0.005$ & $-0.004$ &     & G0V \\ 
HT Vir   &  67186 & 119931 & 0.40767 &   & EX & EW & 7.07 & 7.86 & 0.51 & 1.06 & 0.56 & 0.42 & 15.4 $\pm$ 2.7    & 3.79 & 0.893 & $+0.005$ & $-0.144$ & $-0.042$ & $+0.005$ & SB2 & F8V \\ 
KR Com   &  65069 & 115955 & 0.40797 & H & EX & EW & 7.15 & 7.63 & 0.52 & 0.56 & 0.12 & 0.07 & 13.1 $\pm$ 0.9    & 3.21 & 0.970 & $+0.001$ & $-0.030$ & $-0.001$ & $+0.003$ & SB2 & G0IV \\
V566 Oph &  87860 & 163611 & 0.40965 &   &    & EW & 7.47 &      & 0.41 &      &      & 0.49 & 14.0 $\pm$ 1.1    & 3.18 & 0.864 & $-0.003$ & $-0.164$ & $-0.026$ & $-0.001$ & SB2 & F4V \\ 
VW Pic   &  23196 &  32278 & 0.42576 & H & EX & EA & 6.97 & 7.14 & 0.30 & 0.16 & 9.90 & 0.09 &\phn 6.4 $\pm$ 1.0 & 1.17 & 0.970 & $-0.002$ & $-0.036$ & $-0.009$ & $+0.003$ &     & F3V \\ 
AW UMa   &  56109 & 99946  & 0.43873 &   &    & EW & 6.84 &      & 0.33 &      &      & 0.24 & 15.1 $\pm$ 0.9    & 2.74 & 0.906 & $+0.001$ & $-0.100$ & $-0.002$ & $+0.004$ & SB2 & F1V \\ 
V972 Her & 87958  &  64078 & 0.44309 &   &    & EL & 6.63 &      & 0.39 &      &      & 0.08 & 16.3 $\pm$ 0.6    & 2.68 & 0.972 & $-0.004$ & $-0.030$ & $-0.003$ & $+0.001$ & SB2 & F4V \\ 
V445 Cep & 109191 & 210431 & 0.44878 & H &    & EL & 6.82 &      & 0.12 &      &      & 0.03 &\phn 8.9 $\pm$ 0.5 & 1.58 & 0.978 & $+0.002$ & $-0.013$ & $+0.001$ & $+0.008$ &     & A0: \\
CN Hyi   &  12884 &  17653 & 0.45611 & H &    & EW & 6.57 &      & 0.41 &      &      & 0.27 & 17.2 $\pm$ 0.6    & 2.75 & 0.906 & $-0.001$ & $-0.100$ & $-0.007$ & $+0.001$ &     & F6V \\
MW Vir   &  69828 & 125048 & 0.49308 & H &    & EL & 6.94 &      & 0.23 &      &      & 0.04 & 12.5 $\pm$ 0.4    & 2.42 & 0.985 & $-0.003$ & $-0.017$ & $-0.001$ & $-0.001$ &     & A5IV \\
V867 Ara &  86658 & 160480 & 0.49372 & H & EX & EW & 7.39 & 7.71 & 0.09 & 0.34 & 0.36 & 0.16 &\phn 5.6 $\pm$ 2.6 & 1.45 & 0.945 & $-0.002$ & $-0.048$ & $-0.002$ & $+0.009$ &     & A1/2V \\
KS Peg   & 116611 & 222133 & 0.50210 &   &    & EL & 5.46 &      & 0.00 &      &      & 0.10 & 13.6 $\pm$ 0.8    & 1.13 & 0.960 & $-0.011$ & $-0.036$ & $+0.003$ & $+0.001$ & SB1 & A1Vn \\
AA Cet   &   9258 &  12180 & 0.53617 &   &    & EB & 6.58 & 7.09 & 0.38 & 0.59 & 8.40 & 0.30 &\phn 4.6 $\pm$ 2.4 & 0.41 & 0.916 & $-0.028$ & $-0.101$ & $-0.018$ & $+0.001$ &     & F2: \\
V918 Her &  82253 & 151701 & 0.57481 & H &    & EL & 7.33 &      & 0.25 &      &      & 0.10 &\phn 8.7 $\pm$ 0.9 & 2.02 & 0.962 & $-0.018$ & $-0.039$ & $+0.001$ & $-0.002$ &     & A2: \\
V353 Peg & 116108 & 221380 & 0.58456 & H &    & EL & 7.42 &      & 0.21 &      &      & 0.07 &\phn 5.7 $\pm$ 0.9 & 1.20 & 0.983 & $-0.009$ & $-0.023$ & $-0.005$ & $-0.001$ &     & A2: \\
eps CrA  &  93174 & 175813 & 0.59144 &   &    & EW & 4.76 &      & 0.34 &      &      & 0.26 & 33.4 $\pm$ 0.9    & 2.35 & 0.917 & $-0.007$ & $-0.101$ & $-0.014$ & $-0.004$ & SB2 & F2V \\
RR Cen   &  69779 & 124689 & 0.60569 &   &    & EW & 7.32 &      & 0.34 &      &      & 0.42 &\phn 9.8 $\pm$ 0.9 & 2.22 & 0.880 & $-0.003$ & $-0.148$ & $-0.026$ & $-0.002$ & SB2 & F0V \\
IS CMa   &  30174 &  44524 & 0.61698 & H &    & EW & 6.87 &      & 0.31 &      &      & 0.50 & 10.0 $\pm$ 0.7    & 1.87 & 0.858 & $+0.007$ & $-0.174$ & $-0.030$ & $-0.002$ &     & F3V \\
V851 Ara &  83802 & 154426 & 0.61737 & H &    & EL & 6.86 &      & 0.24 &      &      & 0.05 & 11.2 $\pm$ 0.8    & 2.10 & 0.982 & $-0.006$ & $-0.020$ & $-0.000$ & $-0.001$ &     & A7III \\
LL Vel   &  54060 &  96008 & 0.61924 &   &    & EL & 6.73 &      & 0.30 &      &      & 0.03 & 12.9 $\pm$ 0.6    & 2.25 & 0.989 & $-0.000$ & $-0.014$ & $+0.000$ & $-0.003$ & SB1 & F0V \\
V535 Ara &  86306 & 159441 & 0.62931 &   &    & EW & 7.15 &      & 0.30 &      &      & 0.60 &\phn 8.9 $\pm$ 0.9 & 1.91 & 0.827 & $+0.010$ & $-0.188$ & $-0.037$ & $-0.005$ & SB2:& A8V: \\
S Ant    &  46810 &  82610 & 0.64835 &   &    & EW & 6.29 &      & 0.30 &      &      & 0.51 & 13.3 $\pm$ 0.7    & 1.95 & 0.844 & $+0.001$ & $-0.164$ & $-0.029$ & $+0.004$ & SB1 & A9V: \\
V1010 Oph&  82339 & 151676 & 0.66142 &   &    & EB & 6.06 &      & 0.20 &      &      & 0.80 & 13.5 $\pm$ 0.8    & 1.71 & 0.845 & $-0.059$ & $-0.179$ & $-0.037$ & $+0.013$ & SB1:& A3V \\
V449 Aur &  29108 &  41578 & 0.70365 & H & EX & EA & 7.41 &      & 0.14 &      &      & 0.13 &\phn 4.7 $\pm$ 0.9 & 0.79 & 0.982 & $-0.005$ & $-0.032$ & $-0.012$ & $-0.002$ &     & A0: \\
V2082 Cyg&  95833 & 183752 & 0.71408 & H &    & EL & 6.63 &      & 0.31 &      &      & 0.05 & 11.0 $\pm$ 0.6    & 1.85 & 0.971 & $-0.001$ & $-0.020$ & $-0.000$ & $+0.009$ &     & F0: \\
KP Peg   & 105882 & 204215 & 0.72720 &   &    & EB & 7.07 & 7.29 & 0.06 & 0.22 & 3.50 & 0.19 &\phn 4.4 $\pm$ 1.7 & 0.49 & 0.936 & $-0.025$ & $-0.064$ & $ 0.002$ & $+0.002$ &     & A2V \\
PP Hya   &  49209 &  87130 & 0.75231 & H &    & EL & 6.79 &      & 0.15 &      &      & 0.07 &\phn 9.3 $\pm$ 0.9 & 1.62 & 0.971 & $-0.001$ & $-0.025$ & $-0.002$ & $+0.006$ &     & A3III \\
FO Vir   &  65839 & 117362 & 0.77557 &   &    & EB & 6.51 &      & 0.26 &      &      & 0.31 & 11.8 $\pm$ 0.9    & 1.87 & 0.921 & $-0.032$ & $-0.091$ & $-0.013$ & $+0.002$ & SB1 & A7m \\
V1130 Tau&  17988 &  24133 & 0.79887 & H & EX & EA & 6.56 &      & 0.34 &      &      & 0.38 & 15.4 $\pm$ 0.8    & 2.49 & 0.907 & $+0.005$ & $-0.124$ & $-0.053$ & $-0.004$ &     & F5m \\
V2388 Oph&  87655 & 163151 & 0.80230 &   &    & EW & 6.17 & 6.37 & 0.41 & 0.20 & 0.09 & 0.30 & 14.7 $\pm$ 0.8    & 2.21 & 0.909 & $-0.010$ & $-0.106$ & $-0.015$ & $-0.000$ & SB2 & F5Vn \\
V335 Peg & 112960 & 216417 & 0.81075 & H &    & EL & 7.25 &      & 0.44 &      &      & 0.03 & 16.3 $\pm$ 0.9    & 3.31 & 0.985 & $-0.011$ & $-0.013$ & $ 0.000$ & $+0.002$ &     & F5: \\
TV Pic   &  22370 &  30861 & 0.85199 &   &    & EW & 7.37 &      & 0.14 &      &      & 0.14 &\phn 5.0 $\pm$ 0.6 & 0.88 & 0.929 & $-0.006$ & $-0.050$ & $-0.003$ & $+0.017$ & SB2 & A2V \\
ES Lib   &  74765 & 135681 & 0.88303 &   &    & EB & 7.10 & 7.21 & 0.21 & 0.10 & 0.19 & 0.47 &\phn 7.9 $\pm$ 0.9 & 1.70 & 0.878 & $-0.045$ & $-0.124$ & $-0.014$ & $+0.013$ & SB1:& A3IV \\
IW Per   &  16591 &  21912 & 0.91718 &   &    & EL & 5.77 &      & 0.13 &      &      & 0.05 & 18.3 $\pm$ 0.8    & 2.08 & 0.978 & $-0.001$ & $-0.022$ & $-0.000$ & $+0.001$ &     & A3V \\
DW Boo   &  70621 & 126676 & 0.92444 & H &    & EL & 7.40 &      & 0.01 &      &      & 0.07 &\phn 5.1 $\pm$ 1.1 & 0.92 & 0.971 & $-0.008$ & $-0.027$ & $+0.003$ & $-0.000$ &     & A0: \\
TU Hor   &  16339 &  21981 & 0.93599 &   &    & EL & 5.90 &      & 0.11 &      &      & 0.13 &\phn 8.8 $\pm$ 0.6 & 0.62 & 0.934 & $-0.004$ & $-0.044$ & $-0.003$ & $+0.024$ & SB1 & A1V \\
GK Cep   & 106226 & 205372 & 0.93615 &   & EX & EA & 6.90 &      & 0.08 &      &      & 0.47 &\phn 5.6 $\pm$ 0.5 & 0.64 & 0.882 & $+0.002$ & $-0.142$ & $-0.043$ & $-0.001$ & SB2:& A2Vn \\
HL Dra   &  91052 & 172022 & 0.94428 & H & EX & EA & 7.28 &      & 0.19 &      &      & 0.31 &\phn 6.4 $\pm$ 0.6 & 1.30 & 0.939 & $-0.038$ & $-0.083$ & $-0.028$ & $+0.005$ &     & A5 \\
DX Aqr   & 108797 & 209278 & 0.94501 &   & EX & EA & 6.36 & 6.90 & 0.41 & 0.65 & 3.80 & 0.41 &\phn 6.9 $\pm$ 2.2 & 1.11 & 0.946 & $-0.053$ & $-0.079$ & $-0.043$ & $-0.004$ & SB1 & A2V \\
BB Scl   &   7372 &   9770 & 0.47653 &   & EX & EA & 7.09 &      & 0.90 &      &      & 0.20 & 42.3 $\pm$ 1.5    & 5.22 & 0.962 & $+0.001$ & $-0.051$ & $-0.035$ & $-0.000$ &     & K3V \\
ER Vul   & 103833 & 200391 & 0.69809 &   & EX & EA & 7.31 &      & 0.57 &      &      & 0.13 & 20.1 $\pm$ 0.9    & 3.83 & 0.972 & $-0.004$ & $-0.037$ & $-0.012$ & $+0.003$ & SB2 & G0V \\
\tablecomments{Columns: (1) Name; (2) (3) HIP \& HD; (4) period in
days; (5) H = discovered by
Hipparcos; (6) EX = excluded from the sample; (8) (9) $V^{max}$ as observed
and corrected for the third light; (10) observed $B-V$, uncorrected for
the third light (see text re.\ 44~Boo); (11) $\beta = L_3/L_{12}^{max}$; 
(12) separation in
arcsec;  (13) observed amplitude of light variations in magnitudes,
uncorrected for the third light; (14) Hipparcos parallax in
milliarcsec; (15) $M_V$ derived from the $V^{max}$ and the parallax;
(16) -- (19) the Fourier cosine coefficients, in light units, 
uncorrected for third light; (20) the sine Fourier
coefficient; (21) current status of radial velocity observations;
(22) estimate of the spectral type.
}
\enddata
\end{deluxetable}